\documentclass[10pt,superscriptaddress,nofootinbib,longbibliography]{revtex4-2}

\usepackage[Symbol]{upgreek}
\usepackage{enumitem}
\usepackage{xcolor} 
\usepackage{tensor}
\usepackage{csquotes}
\usepackage{amsmath,amsfonts}
\usepackage{amssymb,latexsym}
\usepackage{color}
\usepackage{graphicx}
\usepackage[normalem]{ulem}

\usepackage{bm}

\usepackage{tikz}
\usetikzlibrary{arrows.meta}

\definecolor{axiscolor}{HTML}{263238}
\definecolor{domainfill}{HTML}{EAF1F5}
\definecolor{mublue}{HTML}{2F6B9A}
\definecolor{varorange}{HTML}{C66A22}
\definecolor{genericteal}{HTML}{2A7F78}
\definecolor{bmorange}{HTML}{C66A22}

\usepackage{hyperref}
\usepackage{booktabs}
\hypersetup{
    colorlinks=true,
    linkcolor=blue,
    filecolor=magenta,      
    urlcolor=cyan,
}

\begin{document}

\title{
Stochastic Processes as Non-Metric Geodesics in Information Geometry
}

\author{T. Koide}
\affiliation{Instituto de F\'{\i}sica, Universidade Federal do Rio de Janeiro, 
21941-972, Rio de Janeiro, RJ, Brazil}
\author{A. van de Venn}
\affiliation{Institute of Physics, University of Tartu, W. Ostwaldi 1, 50411 Tartu, Estonia}

\begin{abstract}
We establish a one-to-one correspondence between geodesics associated with the one-parameter family of $\alpha$-connections on the Gaussian statistical manifold and a class of continuous stochastic processes characterized by a time-independent noise intensity.
We demonstrate that geodesics in expectation parameters naturally classify into three distinct geometric categories, among which the Boundary-connecting class allows us to construct an explicit linear stochastic realization with constant diffusion, representing a generalized bridge process.
This result demonstrates how continuous stochastic processes within this Gaussian class can be extended along geometric curves.
Under appropriate operational limits, this generalized bridge process reduces to fundamental stochastic dynamics, either Ornstein-Uhlenbeck (OU) relaxation or free diffusion.
Crucially, the physical restoring force governing the resulting OU relaxation directly determines the underlying connection parameter $\alpha$, providing a concrete physical observable to constrain the manifold geometry.
Depending on the chosen affine connection representation, this restoring force can be attributed either to curvature or to non-metricity, establishing a direct conceptual analogy with the Geometrical Trinity of Gravity.
Furthermore, applying this framework to driven stochastic thermodynamics, we show that the work-minimizing optimal protocol in the slow-driving limit coincides precisely with an expectation geodesic 
of the statistical manifold with non-metricity equipped with $(g, {}^{(1/2)}\Gamma, {}^{(-1/2)}\Gamma)$, highlighting the active physical role of non-metricity in information geometry.
\end{abstract}

\maketitle

\section{Introduction}

Information geometry is a powerful mathematical framework that introduces a differential geometric structure into the space of probability distributions \cite{amari1985,amari2000,amari2016,ay2017}.
In recent years, this framework has been applied to various fields of physics \cite{brody2009,ItoDechant2020,Kim2021,Ito2024}.
A characteristic feature of information geometry is the natural appearance of so-called dual affine connections.
In general, these connections violate metric compatibility, signifying the presence of non-metricity.
The concept of non-metricity has broader historical roots in theoretical physics.
It was Hermann Weyl who originally introduced non-metricity in his attempt to unify the electromagnetic and gravitational fields \cite{weyl1918english}.
While his specific unified theory was eventually superseded, his geometric insight into local scale invariance deeply influenced the modern understanding of gauge symmetry.
Nowadays, non-metricity appears most prominently in Metric-Affine Gravity, a geometric extension of General Relativity \cite{Hehl95, Hehl99, Baldazzi21, Francois25, Blagojevic12, Obukhov06, Mielke17}.
This active study of non-metricity in gravitational physics stands in stark contrast to the current status of information geometry, where the vast majority of physical applications focus exclusively on the Riemannian substructure and ignore non-metric effects.
For instance, a comprehensive review by Brody and Hook explores the extensive applications of information geometry to thermodynamics \cite{brody2009},
basing their discussions primarily on the assumption of the metric-compatible Levi-Civita (LC) connection.
In the present work, we follow a different route: non-metricity will play a central role.

If the differential geometric formulation of probability spaces is to be regarded as more than a convenient mathematical formalism, its intrinsic geometric constructs, such as geodesics, should embody concrete physical significance.
Our primary objective is to investigate the properties of geodesics on a statistical manifold endowed with non-metricity.
Throughout this work, we use the term geodesic in the standard information-geometric sense, namely to denote an autoparallel curve with respect to a chosen affine connection.
Thus, unless stated otherwise, the geodesics considered here are connection-dependent objects rather than length-extremizing curves determined by the LC connection.
Consequently, the geodesic structure in our information-geometric framework depends explicitly on the non-metricity associated with the chosen connection.
More specifically, we investigate the relation between stochastic processes and geodesics on statistical manifolds.
The interpretation of stochastic processes as curves or geodesics on statistical manifolds has recently attracted increasing attention in machine learning, particularly in the description of generative dynamics in diffusion models through the Fisher metric \cite{Karczewski2025}.

Note, however, that the time evolution generated by a stochastic process need not remain confined to a prescribed finite-dimensional statistical manifold.
For example, Ohara and collaborators analyzed the solutions of the porous medium equation, a nonlinear Fokker-Planck equation, using the methods of information geometry \cite{ohara, oharawada2010}.
Since a generic solution of this equation does not necessarily remain within a finite-dimensional $q$-Gaussian family, they considered its moment-conserving projection onto the manifold of $q$-Gaussian densities and studied the relation between the projected trajectory and the corresponding geodesic structure.
In order to avoid this projection step and to focus on stochastic processes whose dynamics remain strictly constrained to a chosen statistical manifold, we restrict our attention to Gaussian statistical manifolds.
More precisely, we consider cases where the Gaussian family is invariant under the stochastic evolution, so that an initially Gaussian distribution remains Gaussian and the dynamics can be described entirely in terms of the Gaussian parameters.
We then study the geodesics of this manifold and their relation to the corresponding stochastic processes.

In this work, we establish a one-to-one correspondence between geodesics associated with the one-parameter family of $\alpha$-connections on the Gaussian statistical manifold and a class of continuous stochastic processes characterized by a time-independent noise intensity.
Following the dualistic framework of information geometry, we equip the manifold with the dual structure $(g, {}^{(\alpha)}\Gamma, {}^{(-\alpha)}\Gamma)$, where ${}^{(\alpha)}\Gamma$ and ${}^{(-\alpha)}\Gamma$ govern parallel transport in the natural and expectation parameters, respectively.
We demonstrate that the geodesics in expectation parameters naturally classify into three distinct geometric categories.
Focusing on the Boundary-connecting class ($\alpha > -1$), we construct a closed-form stochastic representation of a generalized bridge process within this constant-diffusion class.
Under appropriate operational limits, this generalized bridge process reduces to fundamental stochastic dynamics: either Ornstein-Uhlenbeck (OU) relaxation or free diffusion.
This formulation demonstrates how continuous stochastic processes can be extended and classified along geometric curves.
Crucially, the physical restoring force governing the resulting OU relaxation directly determines the underlying connection parameter $\alpha$, providing a concrete physical observable to uniquely constrain the manifold geometry.
Furthermore, applying this framework to driven stochastic thermodynamics, we show that the work-minimizing optimal protocol in the slow-driving limit coincides precisely with an expectation geodesic on the statistical manifold equipped with $(g, {}^{(1/2)}\Gamma, {}^{(-1/2)}\Gamma)$.
Finally, depending on the chosen affine connection representation, this restoring force can be attributed either to curvature or to non-metricity, establishing a direct conceptual analogy with the Geometrical Trinity of Gravity and highlighting the active physical role of non-metricity in information geometry.

The remainder of this paper is organized as follows. 
Section~\ref{sec:gaussian} introduces the basic concepts of information geometry for the case of exponential distributions and defines the Gaussian statistical manifold. 
Section~\ref{sec:classification} provides exact analytical solutions for the general geodesics in expectation-parameter space and classifies their geometric orbits. 
In Sec.~\ref{sec:sto_process_GGM}, 
we formulate the stochastic processes that correspond to these general geodesics. 
Section~\ref{sec:optimization} applies our geometric approach to stochastic thermodynamics and shows that, for a specific non-metric connection, the optimal protocol coincides with an expectation geodesic. 
The analogy between our findings and the geometric trinity of gravity is discussed in Sec.~\ref{sec:gravity}. Section~\ref{sec:conclusion} summarizes our conclusions and outlines future perspectives.

\section{Information Geometry of Exponential Families}
\label{sec:gaussian}

Consider random variables $x=(x_1,x_2,\cdots)$ that are distributed according to the exponential family. Their corresponding probability distribution function $P(x; \theta)$ is given by
\begin{equation}
P(x; \theta) = \exp \left( \theta^i F_i(x) - \psi(\theta) + k(x) \right) \, .
\end{equation}
The exponential family can be equipped with the structure of a smooth manifold, thereby constituting a statistical manifold. 
In this case, the parameters $\theta = (\theta^1, \theta^2, \ldots, \theta^n)$ represent the so-called natural parameters of the statistical manifold. 
The functions $F_i(x)$, where $i=1,\ldots,n$, are termed sufficient statistics
whereas $\psi(\theta)$ is the potential which normalizes the distribution:
\begin{equation}
\int P(x; \theta) dx = 1 \, .
\end{equation}
Note that the function $k(x)$ is irrelevant in the construction of a statistical manifold.
We apply the Einstein summation convention throughout this paper, where repeated upper and lower indices imply a summation over their range.

In information geometry, there are two fundamental geometrical quantities: the Fisher information metric and Amari-Chentsov's cubic tensor which are, respectively, defined by
\begin{align}
g_{ij}(\theta) &:= \mathbb{E}_\theta [ l_i(x; \theta) l_j(x; \theta) ] 
\, , \\
T_{ijk}(\theta) &:= \mathbb{E}_\theta [ l_i(x; \theta) l_j(x; \theta) l_k(x; \theta) ] \, ,
\end{align}
where $\mathbb{E}_\theta[\cdots]$ denotes the expectation value with respect to $P(x; \theta)$,  
and we employ the score function:
\begin{equation}
l_i(x; \theta) = \frac{\partial}{\partial \theta^i} \ln P(x; \theta) \, .
\end{equation}
For the exponential family, 
these fundamental quantities are expressed through the derivatives of the potential:
\begin{align}
g_{ij}(\theta) &= \frac{\partial^2 \psi(\theta)}{\partial \theta^i \partial \theta^j} \, \label{metrc_exp},\\
T_{ijk}(\theta) &= \frac{\partial^3 \psi(\theta)}{\partial \theta^i \partial \theta^j \partial \theta^k}.\label{conn_exp}
\end{align}

The cubic tensor can be used to define a one-parameter family of affine connections on the statistical manifold. 
These are the $\alpha$-connections which are denoted by ${}^{(\alpha)}\Gamma_{ijk}(\theta)$ and are defined by combining the LC-connection ${}^{(0)}\Gamma_{ijk}(\theta)$ and the cubic tensor $T_{ijk}(\theta)$ in the following manner: 
\begin{equation}
{}^{(\alpha)}\Gamma_{ijk}(\theta) := {}^{(0)}\Gamma_{ijk}(\theta) - \frac{\alpha}{2} T_{ijk}(\theta) \, .
\end{equation}
The LC-connection is the standard metric connection in Riemannian geometry.
Its components are determined by the derivatives of the metric tensor which furthermore happen to  correspond to the cubic tensor through Eqs.\ \eqref{metrc_exp} and \eqref{conn_exp} in the exponential family:
\begin{equation}
{}^{(0)}\Gamma_{ijk}(\theta) 
= \frac{1}{2} \left( \frac{\partial g_{ik}(\theta)}{\partial \theta^j} + \frac{\partial g_{ij}(\theta)}{\partial \theta^k} - \frac{\partial g_{jk}(\theta)}{\partial \theta^i} \right) 
= \frac{1}{2} T_{ijk}(\theta)
\, .
\end{equation}
Substituting this relation into the definition of the $\alpha$-connection, 
the components of the $\alpha$-connection in the natural parameters $\theta$ become proportional to the cubic tensor:
\begin{equation}
{}^{(\alpha)}\Gamma_{ijk}(\theta) = \frac{1-\alpha}{2} T_{ijk}(\theta) \, . \label{eqn:alpha_gamma_T}
\end{equation}
This formula reveals a crucial geometric property of exponential families. 
Upon setting $\alpha = 1$, the connection components vanish completely: ${}^{(1)}\Gamma_{ijk}(\theta) = 0$. 
Hence, for $\alpha = 1$, the underlying statistical manifold is flat.

Since the Fisher metric $g_{ij}(\theta)$ is positive definite, the potential $\psi(\theta)$ is convex due to Eq.\ \eqref{metrc_exp}.
We can thus introduce the dual potential $\phi(\eta)$ through the Legendre transformation:
\begin{equation}
  \phi(\eta) = \max_{\theta}\, \left( \theta^i \eta_i - \psi(\theta) \right) \, .
\end{equation}
The dual coordinates, $\eta_i$, also called the expectation parameters, are defined as the gradients of the potential $\psi(\theta)$ and correspond here to the expectation values of the sufficient statistics $F_i(x)$:
\begin{equation}
  \eta_i := \frac{\partial \psi}{\partial \theta^i}
  =\mathbb{E}_\theta [F_i (x)] \, .
\end{equation}
In a similar fashion as above, we can define the metric and cubic tensor with respect to this dual coordinate as
\begin{align}
g^{ij}(\eta) &:= \mathbb{E}_{\eta} \left[ l^i(x; \eta)l^j(x; \eta)\right] = \frac{\partial^2 \phi(\eta)}{\partial \eta_i \partial \eta_j} \, ,\\
T^{ijk}(\eta) &:= \mathbb{E}_{\eta} \left[ l^i(x; \eta)l^j(x; \eta)l^k(x; \eta) \right] =\frac{\partial^3 \phi(\eta)}{\partial \eta_i \partial \eta_j \partial \eta_k} \, , \label{conn_exp2}
\end{align}
where $l^i(x;\eta) := \partial \ln P(x; \eta) /\partial \eta_i$ with $P(x;\eta) = P(x; \theta(\eta))$.
The metrics then satisfy
\begin{align}
g_{ij} (\theta) g^{jk}(\eta) = \delta^k_i \, . \label{eqn:gg=I}
\end{align}
Furthermore, the dual $\alpha$-connection ${}^{(-\alpha)}\Gamma (\eta)$ in the expectation parameter space is defined by
\begin{align}
{}^{(-\alpha)}\Gamma^{ijk}(\eta) &:= \frac{1}{2} \left( \frac{\partial g^{ik}(\eta)}{\partial \eta_j} + \frac{\partial g^{ij}(\eta)}{\partial \eta_k} - \frac{\partial g^{jk}(\eta)}{\partial \eta_i} \right) - \frac{\alpha}{2} T^{ijk}(\eta) \nonumber \\
&= \frac{1-\alpha}{2} T^{ijk}(\eta)
\, .
\label{eqn:alpha_gamma^*_T}
\end{align}
Note that the first term on the right-hand side of the first line corresponds to the dual LC connection in the sense that ${}^{(0)}\Gamma^{ijk}(\eta) := ({}^{(0)}\nabla_{\partial^j} \partial^k, \partial^i)_g$ with $\partial^i = \partial/\partial \eta_i$ for a curved geometry without torsion or non-metricity. 
Here, $(u, v)_g$ denotes the Riemannian inner product on the tangent space induced by the metric, satisfying $(\partial^i, \partial^j)_g = g^{ij}(\eta)$.

The cubic tensor defined in Eq.~(\ref{conn_exp}) through the derivatives of the potential $\psi(\theta)$ and that defined in Eq.~(\ref{conn_exp2}) through the dual potential $\phi(\eta)$ represent distinct geometric quantities.
Specifically, transforming the coordinate components of Eq.~(\ref{conn_exp}) to expectation coordinates does not reproduce Eq.~(\ref{conn_exp2}) directly; instead, they are related with an explicit minus sign:
\begin{equation}
T^{iab}(\eta) = - g^{il}(\eta) g^{ja}(\eta) g^{kb}(\eta) T_{ljk}(\theta) \, ,
\end{equation}
which can be shown by taking the derivative of Eq.\ (\ref{eqn:gg=I}) with respect to $\theta$.
While one could introduce distinct symbols for these two tensors, a common convention in the information geometry literature is to use the same symbol $T$ and distinguish their definitions by the argument.
In this manuscript, we follow this convention: $T(\theta)$ uniquely denotes the derivatives with respect to the natural parameters $\theta$ of $\psi(\theta)$, whereas $T(\eta)$ denotes the derivatives with respect to the expectation parameters $\eta$ of $\phi(\eta)$.

Let us denote the covariant derivative associated with the $\alpha$-connection in the natural parameters $\theta$ by ${}^{(\alpha)}\nabla$. 
Applying ${}^{(\alpha)}\nabla$ to the metric tensor $g_{ij}$ gives
\begin{equation}
{}^{(\alpha)}\nabla_k g_{ij} (\theta)
= 
\alpha T_{ijk} (\theta)
= {}^{(\alpha)}\tensor{\Gamma}{_{ijk}} (\theta) - {}^{(-\alpha)}\tensor{\Gamma}{_{ijk}}(\theta)
\, .
\label{eqn:non-metricity}
\end{equation}
This shows that in general ${}^{(\alpha)}\nabla g \neq 0$ for a non-zero $\alpha$, meaning that the $\alpha$-connection explicitly breaks the metric compatibility.
Therefore, a statistical manifold with non-metricity in information geometry is characterized by three fundamental quantities: $(g, {}^{(\alpha)}\Gamma, {}^{(-\alpha)}\Gamma)$. Here, the former connection ${}^{(\alpha)}\Gamma$ is associated with the natural parameters, while its dual ${}^{(-\alpha)}\Gamma$ is associated with the expectation parameters.

The Fisher metric and the cubic tensor are closely related to a fundamental quantity in information theory, the Kullback-Leibler (KL) divergence, which quantifies the informational difference between two probability distributions:
\begin{equation}
D_{KL}(A | B) = \mathbb{E}_{\theta_A} \left[ \ln \frac{P(x; \theta_A)}{P(x; \theta_B)} \right] \, .
\end{equation}
When a state $\theta_B = \theta + d\theta$ is infinitesimally close to that by $\theta_A = \theta$,
the KL divergence is expanded as 
\begin{equation}
D_{KL}(A | A+dA) = \frac{1}{2} g_{ij}(\theta) d\theta^i d\theta^j + \frac{1}{6} T_{ijk}(\theta) d\theta^i d\theta^j d\theta^k + O(||d\theta||^4) \, .
\end{equation}
This result clearly shows that the Fisher metric represents the symmetric quadratic distance between infinitesimally close states.
Furthermore, the cubic tensor generates the third-order correction, which introduces the fundamental asymmetry of the KL divergence, $D_{KL}(A | B) \neq D_{KL}(B | A) $.
As shown in Eq.\ (\ref{eqn:non-metricity}), 
the origin of the violation of the metric compatibility is caused by this asymmetry of the KL divergence.

As a final remark, it is worth noting that the KL divergence can be obtained from a double integral of the Fisher metric.
Let us consider that the natural parameters change from $\theta_A$ to $\theta_B$ following a trajectory in the Gaussian statistical manifold which is parametrized by $t$ ($t_A \le t \le t_B$), 
satisfying $\theta_A$ at $t_A$ and $\theta_B$ at $t_B$.
Then the KL divergence for the exponential family is given by the double integral:
\begin{align}
  D_{KL} (A|B) 
  &= \iint_{t_A \le t \le s \le t_B} d\eta_i (t) d\theta^i(s) \nonumber \\
  &= \int^{t_B}_{t_A} dt \int^{t_B}_t ds \, g_{ij}(\theta(t)) \dot{\theta}^i(t) \dot{\theta}^j(s) \, ,
\end{align}
where $\dot{\theta}^i(t) := d {\theta}^i (t)/dt$.  
In the second equality, we used $d\eta_i = g_{ij} d\theta^j$.
This can be shown, for example, implementing the integral for $s$ first:
\begin{equation}
\int_{t_A}^{t_B} (\theta_B - \theta(t))^i \dot{\eta}_i (t) dt 
= 
 \phi(\eta_A) - \phi(\eta_B) - \theta_B^i (\eta_{A} - \eta_{B})_i 
\, ,
\end{equation}
where we used $\phi(\eta_A) - \phi(\eta_B) = -\int_{t_A}^{t_B} \theta^i \dot{\eta}_i dt$.
The right-hand side reduces to the dual Bregman divergence, which is identical to the KL divergence through $D_{KL} (A|B) = D^*_{BR} (\eta_A|\eta_B)$.

\subsection{Gaussian statistical manifold}
As a more concrete example, we consider probability distributions in the form of a generalized Gaussian distribution for $f(x)$:
\begin{align}
P(x; \mu, \sigma^2)
= \frac{1}{\sqrt{2\pi\sigma^2}} \exp\left( -\frac{(f(x)-\mu)^2}{2\sigma^2} \right)\left|\frac{df(x)}{dx}\right| \, ,
\label{eqn:GPD}
\end{align}
where $f(x)$ is a strictly monotonic, differentiable function that maps the domain of the random variable $x$ onto the entire real line.
Such a Gaussian distribution takes the form of the exponential family:
\begin{align}
P(x; \theta) = \exp \left( \sum_{i=1}^2 \theta^i F_i (x) - \psi(\theta) + k(x) \right) \, ,
\label{eqn:pd_gsm}
\end{align}
with $(F_1 (x) ,F_2 (x)) = (f(x), (f(x))^2 )$.
Indeed, a comparison of the two expressions above determines the natural parameters $\theta = (\theta^1, \theta^2)$ as
\begin{align}
\theta^1 &= \frac{\mu}{\sigma^2} \, ,\\
\theta^2 &= -\frac{1}{2\sigma^2} \, ,
\end{align}
and $k(x) = \ln |df(x)/dx|$.
Thus, since $\sigma^2 > 0$,
the Gaussian statistical manifold in $\theta$-coordinates corresponds to the lower half plane $(\theta^1, \theta^2) \in \mathbb{R} \times \mathbb{R}_{<0}$. 
The line $\theta^2 = 0$ represents the limit where the variance diverges towards infinity corresponding to a completely delocalized distribution.
Points on this line lie on the boundary of the Gaussian statistical manifold.

The potential $\psi(\theta)$ is here explicitly found to be
\begin{align}
\psi(\theta) &= \frac{\mu^2}{2\sigma^2} + \frac{1}{2}\ln(2\pi\sigma^2) \nonumber \\
    &= -\frac{(\theta^1)^2}{4\theta^2} + \frac{1}{2}\ln(\pi) - \frac{1}{2}\ln(-\theta^2) \, .
    \label{eqn:gsm_potential}
\end{align}
Hence the Fisher metric is given by
\begin{equation}
g_{ij}(\theta) 
= \left( 
\begin{array}{cc}
-\frac{1}{2\theta^2} & \frac{\theta^1}{2(\theta^2)^2} \\
\frac{\theta^1}{2(\theta^2)^2} & \frac{\theta^2 -  (\theta^1)^2}{2(\theta^2)^3}
\end{array}
\right) 
= \left( 
\begin{array}{cc}
\sigma^2 & 2\mu \sigma^2 \\
2\mu \sigma^2 & 4 \mu^2 \sigma^2 + 2 \sigma^4
\end{array}
\right) \, .
\label{eqn:F_metric_theta}
\end{equation}

The expectation parameters $\eta_i$ are computed as
\begin{align}
    \eta_1 &= \frac{\partial \psi}{\partial \theta^1} = -\frac{\theta^1}{2\theta^2} = \mathbb{E}_\theta [f(x)] = \mu \, ,\\
    \eta_2 &= \frac{\partial \psi}{\partial \theta^2} = \frac{(\theta^1)^2}{4(\theta^2)^2} - \frac{1}{2\theta^2} = \mathbb{E}_\theta [(f(x))^2] = \mu^2 + \sigma^2 \, .
\end{align}
In $\eta$-coordinates, the Gaussian statistical manifold corresponds to the region inside the parabola defined by $\eta_2 = \eta_1^2$. The parabola itself provides the boundary corresponding to the limit of zero variance.
Furthermore, the dual potential $\phi(\eta)$ is found to be
\begin{align}
\phi(\eta) &= - \frac{1}{2} - \frac{1}{2}\ln(2\pi\sigma^2) \, , \\
 &= - \frac{1}{2} \ln(\eta_2 - \eta_1^2) - \frac{1}{2} \left( 1 + \ln(2\pi) \right).
\end{align}
Note that, disregarding the irrelevant constant term, $\phi(\eta)$ represents the negative Shannon entropy.
Finally, the dual Fisher metric is readily computed as
\begin{equation}
   g^{ij}(\eta) 
= \left( 
\begin{array}{cc}
\frac{\eta_2 + (\eta_1)^2}{(\eta_2 - (\eta_1)^2)^2} & - \frac{\eta_1}{(\eta_2 - (\eta_1)^2)^2} \\
- \frac{\eta_1}{(\eta_2 - (\eta_1)^2)^2} & \frac{1}{2(\eta_2 - (\eta_1)^2)^2}
\end{array}
\right) 
= \left( 
\begin{array}{cc}
\frac{2\mu^2 + \sigma^2}{\sigma^4} & -\frac{\mu}{\sigma^4} \\
-\frac{\mu}{\sigma^4} & \frac{1}{2\sigma^4}
\end{array}
\right) \, . \label{eqn:g^uu}
\end{equation}

\section{Geometric Classification of General $\alpha$-Geodesics}
\label{sec:classification}

As seen previously, a statistical manifold is characterized by $(g, {}^{(\alpha)}\Gamma, {}^{(-\alpha)}\Gamma)$. Thus its geometric structure depends on the parameter $\alpha \in \mathbb{R}$.
In the study of the exponential families, 
one typically considers the case $\alpha=1$.
The geodesics associated with ${}^{(e)}\Gamma (\theta):={}^{(1)}\Gamma(\theta)$ are called the e-geodesics (exponential geodesics), while the geodesics associated with ${}^{(m)}\Gamma(\eta):={}^{(-1)}\Gamma(\eta)$ are called the m-geodesics (mixture geodesics).
The natural parameters constitute an affine coordinate system for the exponential connection, meaning that ${}^{(e)}\Gamma_{ijk}(\theta) =0$, 
as shown in Eq.\ (\ref{eqn:alpha_gamma_T}).
Similarly,
as shown in Eq.\ (\ref{eqn:alpha_gamma^*_T}), the expectation parameters constitute an affine coordinate system for the mixture connection since ${}^{(-1)}\Gamma^{ijk}(\eta) =0$.
This appearance of dual affine coordinates is known as 
a dually flat structure.

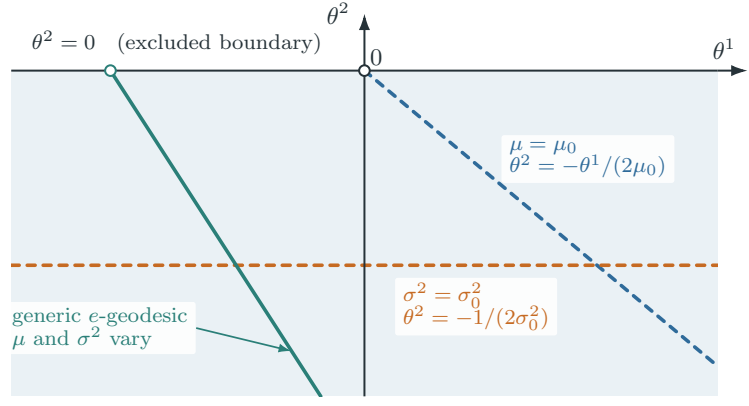
\begin{figure}
\begin{center}
\begin{tikzpicture}[
    x=1.05cm,
    y=1.05cm,
    font=\small,
    axis/.style={
        axiscolor,
        line width=0.75pt,
        -{Latex[length=2.2mm,width=1.5mm]}
    },
    boundary/.style={
        axiscolor,
        line width=0.75pt
    },
    geodesic/.style={
        line width=1.35pt,
        line cap=round
    },
    curve label/.style={
        fill=white,
        fill opacity=0.90,
        text opacity=1,
        inner xsep=3pt,
        inner ysep=2pt,
        rounded corners=1pt,
        align=left,
        font=\footnotesize
    }
]

\def\xmin{-4.45}
\def\xmax{ 4.45}
\def\ymin{-4.10}

\path[fill=domainfill]
    (\xmin,\ymin) rectangle (\xmax,0);


\begin{scope}
    \clip (\xmin,\ymin) rectangle (\xmax,-0.015);

    \draw[geodesic, mublue, dashed]
        (0,0) -- (4.45,-3.72);

    \draw[geodesic, varorange, dashed]
        (\xmin,-2.45) -- (\xmax,-2.45);

    \draw[geodesic, genericteal]
        (-3.20,0) -- (-0.55,\ymin);
\end{scope}

\node[curve label, text=mublue, anchor=south west]
    at (1.72,-1.42)
    {$\mu=\mu_0$\\[-1pt]
     $\theta^2=-\theta^1/(2\mu_0)$};

\node[curve label, text=varorange, anchor=south west]
    at (0.38,-3.34)
    {$\sigma^2=\sigma_0^2$\\[-1pt]
     $\theta^2=-1/(2\sigma_0^2)$};

\node[curve label, text=genericteal, anchor=east]
    (generic-label) at (-2.12,-3.24)
    {generic $e$-geodesic\\[-1pt]
     $\mu$ and $\sigma^2$ vary};
\draw[genericteal, line width=0.65pt,
      -{Latex[length=1.7mm,width=1.2mm]}]
    (generic-label.east) -- (-0.92,-3.53);

\draw[boundary]
    (\xmin,0) -- (\xmax,0);
\draw[axis]
    (\xmax,0) -- (\xmax+0.38,0)
    node[above left=1pt, axiscolor] {$\theta^1$};
\draw[axis]
    (0,\ymin) -- (0,0.72)
    node[left=2pt, axiscolor] {$\theta^2$};

\node[axiscolor, anchor=south west, font=\footnotesize]
    at (\xmin+0.15,0.10)
    {$\theta^2=0$ \; (excluded boundary)};
\filldraw[fill=white, draw=axiscolor, line width=0.75pt]
    (0,0) circle[radius=2.0pt];
\node[axiscolor, anchor=south west, inner sep=2pt]
    at (0,0) {$0$};

\filldraw[fill=white, draw=genericteal, line width=0.75pt]
    (-3.2,0) circle[radius=2.0pt];

\end{tikzpicture}
\end{center}
\caption{
The Gaussian manifold is shown in the $(\theta^1, \theta^2)$ plane.
The physically permissible states of this manifold lie strictly in the region below $\theta^2 = 0$.
An e-geodesic corresponds to an arbitrary straight line in this $\theta$-plane.
For example, a straight line passing through the origin represents a specific process where the mean $\mu$ remains constant.
A straight line parallel to the $\theta^1$-axis represents a process where the variance $\sigma^2$ remains constant.
A general straight line not passing through the origin describes a process where both $\mu$ and $\sigma^2$ change.
}
\label{fig:tra1}
\end{figure}

The e-geodesics of the Gaussian statistical manifold are shown in Fig.\ \ref{fig:tra1}.
Since $\theta^2 = -1/(2\sigma^2) < 0$, 
this manifold corresponds to the lower half of the $(\theta^1, \theta^2)$-plane.
Due to the vanishing of the exponential connection, an e-geodesic
corresponds to
a straight line in the lower half $(\theta^1, \theta^2)$-plane.
For example, a straight line directed toward the origin (excluding the origin itself), which is realized when $\theta^1(t) \propto \theta^2(t)$, represents a specific process where the mean $\mu$ remains constant.
A general straight line not passing through the origin describes a process where both $\mu$ and $\sigma^2$ change. 
A straight line parallel to the $\theta^1$-axis represents a process where the variance $\sigma^2$ remains constant.

\begin{figure}
\begin{center}
\begin{tikzpicture}[
    x=1.08cm,
    y=0.92cm,
    font=\small,
    axis/.style={
        axiscolor,
        line width=0.75pt,
        -{Latex[length=2.2mm,width=1.5mm]}
    },
    boundary/.style={
        axiscolor,
        line width=1.05pt,
        line cap=round
    },
    geodesic/.style={
        genericteal,
        line width=1.25pt,
        line cap=round
    },
    brownian/.style={
        bmorange,
        line width=1.25pt,
        line cap=round    
    },
    curve label/.style={
        fill=white,
        fill opacity=0.92,
        text opacity=1,
        inner xsep=3pt,
        inner ysep=2pt,
        rounded corners=1pt,
        align=left,
        font=\footnotesize
    },
    open m point/.style={
        circle,
        draw=genericteal,
        fill=white,
        line width=0.85pt,
        inner sep=1.55pt
    },
    open bm point/.style={
        circle,
        draw=bmorange,
        fill=white,
        line width=0.85pt,
        inner sep=1.55pt
    }
]

\def\xleft{-2.32}
\def\xright{2.32}
\def\ytop{5.40}

\path[fill=domainfill]
    plot[domain=\xleft:\xright, samples=120, smooth]
        (\x,{\x*\x})
    -- (\xright,\ytop)
    -- (\xleft,\ytop)
    -- cycle;

\draw[axis]
    (-2.65,0) -- (2.72,0)
    node[above left=1pt, axiscolor] {$\eta_1$};
\draw[axis]
    (0,0) -- (0,5.70)
    node[left=2pt, axiscolor] {$\eta_2$};

\coordinate (A) at (-1.75,{(-1.75)*(-1.75)});
\coordinate (B) at ( 1.20,{( 1.20)*( 1.20)});
\coordinate (C) at (-0.62,{(-0.62)*(-0.62)});
\coordinate (D) at ( 2.12,{( 2.12)*( 2.12)});

\draw[geodesic] (A) -- (B);
\draw[geodesic] (C) -- (D);

\coordinate (BM0) at (0.58,{0.58*0.58});
\draw[brownian,dashed] (BM0) -- (0.58,5.38);

\draw[boundary]
    plot[domain=\xleft:\xright, samples=120, smooth]
        (\x,{\x*\x});

\node[open m point] at (A) {};
\node[open m point] at (B) {};
\node[open m point] at (C) {};
\node[open m point] at (D) {};
\node[open bm point] at (BM0) {};


\node[curve label, text=genericteal, anchor=east]
    (m-label) at (-1.83,1.27)
    {$m$-geodesic\\[-1pt]
     (bridge process)};
\draw[mublue, line width=0.65pt,
      -{Latex[length=1.7mm,width=1.2mm]}]
    (m-label.north east) -- (-1.35,2.7);
\draw[genericteal, line width=0.65pt,
      -{Latex[length=1.7mm,width=1.2mm]}]
    (m-label.south east) -- (-0.55,0.7);

\node[curve label, text=bmorange, anchor=south west]
    at (0.78,5.40)
    {standard Brownian motion\\[-1pt]
     $\eta_1=\mu_A$};

\node[curve label, text=axiscolor, anchor=west]
    (boundary-label) at (1.62,1.13)
    {$\eta_2=(\eta_1)^2$\\[-1pt]
     (excluded boundary)};
\draw[axiscolor, line width=0.60pt,
      -{Latex[length=1.6mm,width=1.1mm]}]
    (boundary-label.north west) + (0.5,0) -- (1.70,2.4);

\node[axiscolor, anchor=north, inner sep=2pt]
    at (0,0) {$0$};

\end{tikzpicture}
\end{center}
\caption{
The Gaussian manifold is shown in the $(\eta_1, \eta_2)$ plane.
The physically permissible states of this manifold lie strictly inside the region bounded by the curve $\eta_2 = (\eta_1)^2$.
An m-geodesic is shown as a part of a straight line connecting two boundary points on the curve.
In contrast, standard Brownian motion, which corresponds to the limit of an m-geodesic with an infinite slope, can only evolve in the direction of increasing $\eta_2$.
The bridge-process interpretation applies specifically to the subclass of maximal non-vertical straight geodesics extended to the deterministic boundary ($\sigma^2 = 0$), pinned at fixed boundary times $t_A$ and $t_B$.
}
\label{fig:tra2}
\end{figure}
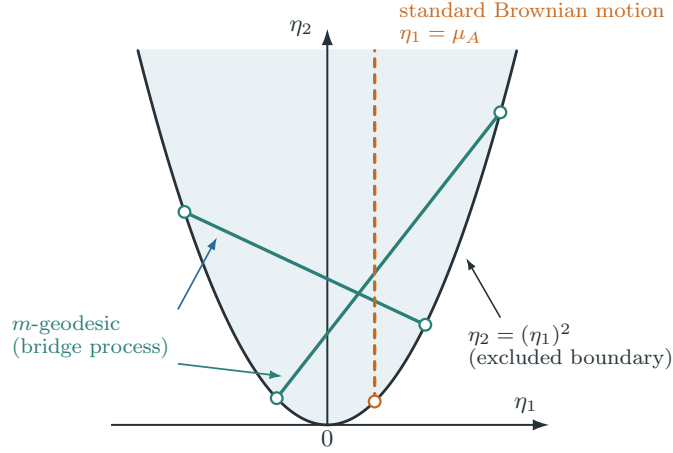

On the other hand, as shown in Fig.\  \ref{fig:tra2}, a straight line within the region bounded by the parabola $\eta_2 = (\eta_1)^2$ in the $\eta$-plane corresponds to an $m$-geodesic.
Although general $m$-geodesics are arbitrary straight segments confined to the interior of the parameter space ($\sigma^2 > 0$), they can be extended until their endpoints touch the boundary $\eta_2 = (\eta_1)^2$.
This boundary corresponds to the limit where the variance vanishes ($\sigma^2 \to 0$), forming the deterministic boundary of the Gaussian statistical manifold.
Since the variance vanishes on this boundary, the probability distribution reduces to a Dirac delta function, meaning the physical state is localized at deterministic values without uncertainty.

In the theory of stochastic processes, a bridge process is defined as a process conditioned to start and end at specific deterministic values at given initial and final times \cite{revuz1999continuous}.
Therefore, the bridge-process interpretation applies specifically to the restricted subclass of maximal, non-vertical $m$-geodesics that extend all the way to this deterministic boundary ($\sigma^2 = 0$), such as the well-known pinned Brownian bridge.
For any Boundary-connecting geodesic ($\alpha > -1$), any non-vertical interior segment ($\sigma^2 > 0$) can be uniquely extended to touch the deterministic boundary ($\sigma^2 = 0$) at two distinct points, meaning that the associated Gaussian process can always be interpreted as a sub-interval of a corresponding pinned bridge process.
Moreover, standard Brownian motion corresponds to the limit of an $m$-geodesic with an infinite slope (a vertical line in the expectation space).
These properties are demonstrated in detail in the next section.

As mentioned, $e$- and $m$-geodesics occupy a special place in the study of information geometry and have been thoroughly investigated. 
In this work, however, we return to the general statistical manifold equipped with $(g, {}^{(\alpha)}\Gamma, {}^{(-\alpha)}\Gamma)$ without restricting ourselves to $\alpha = 1$.
Then, for the Gaussian statistical manifold, the general geodesic equation in expectation parameters is given by
\begin{equation}
{}^{(-\alpha)}\nabla_{(\dot{\eta}_i \partial^i)} (\dot{\eta}_k \partial^k) = 0 
\, \longrightarrow \, 
g^{ij} \ddot{\eta}_j + \frac{1-\alpha}{2} T^{ijk} (\eta)\dot{\eta}_j \dot{\eta}_k = 0 \, ,
\end{equation}
leading to
\begin{align}
\ddot{\mu} - \frac{1-\alpha}{\sigma^2}\dot{\mu} \frac{d \sigma^2}{dt}  &= 0 \, , \label{eqn:eta_geo_mu}\\
\frac{d^2 \sigma^2}{dt^2} + (1+\alpha)\dot{\mu}^2 - \frac{1-\alpha}{\sigma^2} \left( \frac{d \sigma^2}{dt}  \right)^2 &= 0 \, , \label{eqn:eta_geo_sigma}
\end{align}
where $\partial^i := \partial/\partial \eta_i$.
Note that the cubic tensor can be evaluated by substituting Eq.\ \eqref{eqn:g^uu} into the relation $T^{ijk} (\eta)= \partial g^{ij} (\eta)/\partial \eta_k$.
From these equations, we obtain
\begin{equation}
\frac{d}{dt} \left[ \left( \frac{d\sigma^2}{dt} \right)^2 (\sigma^2)^{2\alpha -2} \right] = -2(1+\alpha) A^2 d\sigma^2/dt \, ,
\label{diff_eq}
\end{equation}
where we used the first equation in the form
\begin{equation}
 \dot{\mu} = A (\sigma^2)^{1-\alpha} \, ,
\label{eqn:mu_dot_A}
\end{equation}
with an integration constant $A$.
The differential equation \eqref{diff_eq} is then integrated as
\begin{equation}
\left( \frac{d\sigma^2}{dt} \right)^2 = (\sigma^2)^{2-2\alpha} \left[ C - 2(1+\alpha) A^2 \sigma^2 \right] \, , \label{eqn:sigma_dot_sq}
\end{equation}
where $C$ is another integration constant.
To find the geodesic trajectories, we express $\mu$ as a function of $\sigma^2$, and hence $d\mu/d\sigma^2 = \dot{\mu}/(d\sigma^2/dt)$. We have, from the above two equations,
\begin{align}
    \frac{d\mu}{d\sigma^2} 
    = \frac{\dot{\mu}}{d\sigma^2/dt} = \pm \frac{A}{\sqrt{C - 2(1+\alpha) A^2 \sigma^2}} \, , \label{eqn:dmu/dsigma2}
\end{align}
which can be analytically integrated over $\sigma^2$. 
Accordingly, any geodesics of ${}^{(-\alpha)} \Gamma (\eta)$ 
in expectation parameters are classified into three categories: Boundary-connecting ($\alpha > -1$), Critical-boundary ($\alpha = -1$) and Non-boundary-connecting ($\alpha < -1$) on the $(\mu, \sigma)$-plane.

Although the explicit mathematical representation of the connection coefficients and the geodesic equations depends on the choice of coordinates ($\theta$ or $\eta$), the resulting geometric trajectories of the geodesics are determined solely by the chosen connection, representing coordinate-independent curves on the statistical manifold. 
As we will discuss in Sec.\ \ref{sec:CooTrans}, the geodesic equations for the expectation parameters can be mapped to the equations for the natural parameters through the coordinate transformation, and they coincide with the geodesics of the natural parameters by changing the parameter $\alpha$.

\subsection{Boundary-connecting geodesics ($\alpha > -1$)} 
\label{sec:elliptical}

When $\alpha > -1$, the integral of the expectation geodesic yields
\begin{equation}
\mu = \mu_0 \mp \frac{1}{(1+\alpha)A} \sqrt{C - 2(1+\alpha)A^2 \sigma^2} \, ,
\label{eqn:eq_mu_alpha>-1}
\end{equation}
where $\mu_0$ is an integration constant. 
By rearranging this, we obtain the equation of an ellipse:
\begin{equation}
(\mu - \mu_0)^2 + \frac{2}{1+\alpha} \sigma^2 = R^2 \, , \label{eqn:ellipse}
\end{equation}
where $R^2 = C / \left( (1+\alpha)^2 A^2 \right)$.

In Section~\ref{sec:sto_process_GGM}, we will discuss the stochastic processes of this case in details. For this purpose, it is convenient to parameterize the geodesics using an angle $\chi$:
\begin{align}
\mu(\chi) &= \mu_0 + R \sin\chi \, , \label{eqn:geo_evolution_mu}\\
\sigma^2(\chi) &= \frac{1+\alpha}{2} R^2 \cos^2\chi \label{eqn:geo_evolution_sigma2}\,,
\end{align}
with $ -\pi/2 \le \chi \le \pi/2$.
By substituting these into Eq.\ (\ref{eqn:mu_dot_A}), we find the dynamical equation for the angle $\chi$:
\begin{equation}
\dot{\chi} = \omega (\cos\chi)^{1-2\alpha} \, ,
\label{eqn:dif_eq_phi}
\end{equation}
where $\omega := A \left( (1+\alpha)/2 \right)^{1-\alpha} R^{1-2\alpha}$ is a constant.
This equation determines the time evolution of the angle $\chi$.

In this case, the geodesic intersects the boundary curve $\eta_2 = (\eta_1)^2$ 
at $\chi = \pm \pi/2$. 
We consider the evolution in the interval $t_A \le t \le t_B$, satisfying $\sigma^2(t_A) = \sigma^2(t_B) = 0$. 
The physical boundary values for the mean are given by $\mu(t_A) = \mu_A$ and $\mu(t_B) = \mu_B$.
Without loss of generality, we assume $\mu_B > \mu_A$ (the case $\mu_B < \mu_A$ is obtained symmetrically by reversing the sign of $R$ or $\omega$).
To satisfy these conditions, the angle parameter evolves from $\chi(t_A) = -\pi/2$ to $\chi(t_B) = \pi/2$.
Thus, we define $\mu_0$ and $R$ as
\begin{equation}
\mu_0 = \frac{\mu_A + \mu_B}{2} \, , \quad R = \frac{\mu_B - \mu_A}{2} \, .
\end{equation}
For $\alpha \neq 1/2$, substituting Eq.~(\ref{eqn:dif_eq_phi}) into Eqs.~(\ref{eqn:geo_evolution_mu}) and (\ref{eqn:geo_evolution_sigma2}) yields the time-dependent mean and variance in terms of $\dot{\chi}(t)/\omega$:
\begin{align}
\mu(t) &= \mu_0 \pm R \sqrt{1 - \left( \frac{\dot{\chi}(t)}{\omega} \right)^{\frac{2}{1-2\alpha}}} \, , \label{eq:mu_dot} \\
\sigma^2(t) &= \frac{1+\alpha}{2} R^2 \left( \frac{\dot{\chi}(t)}{\omega} \right)^{\frac{2}{1-2\alpha}} \, . \label{eq:sigma_dot}
\end{align}
For $\alpha = 1/2$, Eq.~(\ref{eqn:dif_eq_phi}) reduces to $\dot{\chi} = \omega$, which straightforwardly integrates to $\chi(t) = \omega(t - t_0)$ with an integration constant $t_0$, yielding
\begin{align}
\mu(t) &= \mu_0 + R \sin ( \omega(t-t_0)) \, , \\
\sigma^2(t) &= \frac{3}{4} R^2 \cos^2 ( \omega(t-t_0) ) \, .
\end{align}

\subsection{Critical-boundary Geodesics ($\alpha = -1$)}

When $\alpha = -1$, the right-hand side of the total derivative equation vanishes. 
From Eq.~(\ref{eqn:dmu/dsigma2}), we find a linear relation in the $(\mu, \sigma^2)$ plane:
\begin{equation}
\sigma^2 \propto (\mu - \mu_0) \,,
\end{equation}
where $\mu_0$ is an integration constant. 
One can easily confirm that this linear relation corresponds to a parabola in the expectation parameter space $(\eta_1, \eta_2) = (\mu, \mu^2 + \sigma^2)$, because $\eta_2 = \eta_1^2 + \sigma^2 \propto \eta_1^2 + c(\eta_1 - \mu_0)$. 
The trajectory that forms a straight line in the expectation parameter space is not this case, but rather the $m$-geodesic corresponding to $\alpha = 1$, for which ${}^{(-1)}\Gamma(\eta) = 0$.

\subsection{Non-boundary-connecting Geodesics ($\alpha < -1$)}

When $\alpha < -1$, the term $(1+\alpha)$ becomes negative. 
The integral of the the expectation geodesic trajectory yields
\begin{equation}
\mu = \mu_0 \pm \frac{1}{(-1-\alpha)A} \sqrt{C + 2(-1-\alpha)A^2 \sigma^2} \, ,
\end{equation}
which corresponds to Eq.\ (\ref{eqn:eq_mu_alpha>-1}) in the case of $\alpha>-1$.
By squaring both sides and rearranging, we obtain the general equation of a hyperbola on the $(\mu, \sigma)$-plane:
\begin{equation}
(\mu - \mu_0)^2 - \frac{2}{(-1-\alpha)} \sigma^2 = \frac{C}{(1+\alpha)^2 A^2} \, . 
\end{equation}
The equation is similar in form to Eq.\ \eqref{eqn:ellipse}, but the sign of the second term on the left-hand side is now negative, making it the defining equation of a hyperbola.
Depending on the sign of the integration constant $C$, this equation describes either a vertically or horizontally opening hyperbola. 
Which case appears in practice is determined by the boundary conditions of the initial and final states.

\subsubsection*{Case 1: Vertically opening hyperbola ($C < 0$)}

When the constant is negative ($C/(1+\alpha)^2 A^2 = -R^2$), the geodesic forms a vertically opening hyperbola:
\begin{equation}
\frac{2}{-1-\alpha} \sigma^2 - (\mu - \mu_0)^2 = R^2 \, .
\end{equation}
This trajectory corresponds to the upper branch ($\sigma^2 > 0$) of the hyperbola, strictly confined within the interior of the Gaussian statistical manifold, ensuring a strictly positive minimum variance.
We parameterize this trajectory using an angle $\chi$:
\begin{align}
\mu(\chi) &= \mu_0 + R \sinh\chi \, , \\
\sigma^2(\chi) &= \frac{-1-\alpha}{2} R^2 \cosh^2\chi \, .
\end{align}
The relation between $\chi$ and $t$ is determined by substituting these into Eq.\ (\ref{eqn:mu_dot_A}):
\begin{equation}
   \dot{\chi} = \omega (\cosh \chi)^{1-2\alpha} \, ,
   \label{eqn:dif_eq_psi_vert}
\end{equation}
where $\omega := A \left((-1-\alpha)/2 \right)^{1-\alpha} R^{1-2\alpha}$.

\subsubsection*{Case 2: Horizontally opening hyperbola ($C > 0$)}

When the constant is positive ($C/(1+\alpha)^2 A^2 = R^2$), 
the geodesic forms a horizontally opening hyperbola:
\begin{equation}
(\mu - \mu_0)^2 - \frac{2}{-1-\alpha} \sigma^2 = R^2 \, .
\end{equation}
The trajectory consists of two disconnected branches ($\mu-\mu_0 >0$ and $<0$). 
For the right branch, the parameterization of $\mu$ is given by
\begin{align}
\mu(\chi) = \mu_0 + R \cosh\chi \, , 
\end{align}
and for the left branch, we have 
\begin{align}
\mu(\chi) = \mu_0 - R \cosh\chi \, .
\end{align}
For both branches, $\sigma^2$ is parametrized by 
\begin{align}
\sigma^2(\chi) &= \frac{-1-\alpha}{2} R^2 \sinh^2\chi \, .
\end{align}
By substituting into Eq.\ (\ref{eqn:mu_dot_A}), we obtain the differential equation for $\chi$:
\begin{equation}
   \dot{\chi} = \omega (\sinh \chi)^{1-2\alpha} \, ,
   \label{eqn:dif_eq_psi_horiz}
\end{equation}
with the same constant $\omega$. 
Crucially, at $\chi \to 0$, this trajectory reaches the singular boundary where the variance vanishes ($\sigma^2 \to 0$). 
Since the vanishing variance pins the probability distribution to a Dirac delta function at $\chi \to 0$, this trajectory explicitly describes a stochastic process that is deterministically pinned at only one temporal boundary (either initially or terminally), in sharp contrast to the two-point bridge processes.

To express $\mu$ and $\sigma$ as functions of $t$ for both cases, we should solve the respective differential equations for $\chi$. 
To solve these equations analytically, we consider specific values of $\alpha$ in the following discussions.

\section{Stochastic Processes Associated with Geodesics on Gaussian Manifolds}
\label{sec:sto_process_GGM}

In the preceding section, we analyzed the geometric classification of $\alpha$-geodesics in terms of the expectation parameters on the Gaussian statistical manifold. 
In this section, we establish the explicit correspondence between these geodesics and continuous stochastic processes. 
Specifically, we formulate linear stochastic differential equations (SDEs) whose time-dependent probability distributions remain strictly Gaussian while tracing out these exact geodesic curves.

Stochastic processes that maintain a Gaussian form throughout their time evolution, given some mean $\mu(t)$ and variance $\sigma^2(t)$, can be realized by the following linear stochastic differential equation (SDE):
\begin{equation}
d X(t) = \left[ \dot{\mu}(t) + \frac{d \sigma^2(t)/dt - 2D}{2\sigma^2(t)} (X(t) - \mu(t)) \right] dt + \sqrt{2D} dW_t \, ,
\label{eqn:sde_general}
\end{equation}
where $D$ characterizes the intensity of the noise term, and $W_t$ represents the standard Wiener process.
In the discretized time representation with a microscopic step width $dt$, the noise increments satisfy
\begin{align}
 \mathbb{E}_{W} [dW_t] &= 0 \, ,\\
 \mathbb{E}_{W} [dW_t \, dW_{t^\prime} ] &= dt \, \delta_{t,t^\prime} \, ,
\end{align}
where $\mathbb{E}_{W}[\cdots]$ denotes the ensemble average for the Wiener process, and $\delta_{t, t'}$ is the Kronecker delta on the discretized time grid.
It is easy to see that the probability distribution (\ref{eqn:GPD}) satisfies the Fokker-Planck equation of the above SDE by choosing $f(x) = x$ \cite{Gardiner_book}. 
Thus, by substituting the solutions $\mu(t)$ and $\sigma^2(t)$ of the geodesic equations into the SDE above, we can explicitly define a specific realization of the stochastic process corresponding to a given geodesic on the statistical manifold.

In the following calculation, we explicitly evaluate the stochastic process for the Boundary-connecting case ($\alpha > -1$). 
In this case, the geodesic necessarily intersects the boundary curve $\eta_2 = (\eta_1)^2$ at two points. Consequently, the corresponding stochastic process can be regarded as a type of bridge process.
To solve this SDE over the time interval $t_A \le t \le t_B$, 
we introduce the following quantities:
\begin{align}
Y(t) &:= X(t) - \mu (t) \, ,\\
\Phi(t;t_A) &:= \exp \left( \int_{t_A}^t \frac{d\sigma^2(u)/du - 2D}{2\sigma^2(u)} du \right) 
   = \frac{\sigma(t)}{\sigma(t_A)} \exp \left( - \int_{t_A}^t \frac{D}{\sigma^2(u)} du \right) \, .
\end{align}
We note that $\Phi(t;t_A)$ remains finite even for a bridge process where $\sigma^2(t_A) = 0$. 
See App.~\ref{app:finite} for details.
To cast the SDE into a more tractable form, we introduce an auxiliary function $G(t)$ defined by
\begin{equation}
    Y(t) = \Phi(t;t_A) G(t) \, .
\end{equation}
The differential equation for $G(t)$ then reduces to a purely noise-driven form:
\begin{equation}
    dG(t) = \frac{\sqrt{2D}}{\Phi (t;t_A)} dW_t \, ,
\end{equation}
satisfying 
\begin{equation}
\tau (t;t_A) :=\int^t_{t_A} (dG (u))^2  = \int^t_{t_A} \frac{2D}{\Phi^2 (u;t_A)} du \, .   
\end{equation}
Thus, $G(t)$ coincides with a time-scaled Wiener process:
\begin{align}
   G(t) &= W_{\tau (t;t_A)} \, ,
\end{align}
and the solution of Eq.~(\ref{eqn:sde_general}) is given by
\begin{equation}
   X(t) = \mu(t) + \Phi(t;t_A) W_{\tau (t;t_A)} \, . \label{eqn:exact_sol_BridgeSDE}
\end{equation}
Substituting $\mu(t)$ and $\sigma^2(t)$ given by Eqs.~(\ref{eqn:geo_evolution_mu}) and (\ref{eqn:geo_evolution_sigma2}) automatically yields the stochastic process for the Boundary-connecting geodesics ($\alpha > -1$):
\begin{align}
X(t) = \mu_0 + R \sin \chi(t) + \sqrt{2D} \cos \chi(t) \int_{t_A}^t \frac{1}{\cos \chi(s)} \exp \left( -\mathcal{K}_\alpha \int_{\chi(s)}^{\chi(t)} \frac{d\chi'}{(\cos \chi')^{3-2\alpha}} \right) dW_s \, ,
\label{eq:general_X_t_main}
\end{align}
where 
\begin{equation}
    \mathcal{K}_\alpha \equiv \frac{2D}{(1+\alpha)R^2 \omega} \, .
\end{equation}
This stochastic process provides an information-geometric generalization of the bridge process.

The formal solution (\ref{eqn:exact_sol_BridgeSDE}) is regular on the half-open interval $[t_A, t_B)$.
The terminal behavior at $t \to t_B$ ($\chi \to \pi/2$) can be evaluated directly by transforming the integral from physical time $t$ to the angle coordinate $\chi$.
Using $dt = d\chi / [\omega (\cos\chi)^{1-2\alpha}]$ and Eq.~(\ref{eqn:geo_evolution_sigma2}), the integral in the propagator takes the explicit form:
\begin{equation}
\int_{t_A}^t \frac{ds}{\sigma^2(s)} = \frac{2}{(1+\alpha)R^2\omega} \int_{-\pi/2}^{\chi(t)} (\cos\chi')^{2\alpha - 3} d\chi' \, .
\end{equation}
For $\alpha = 1$, this directly yields the logarithmic divergence $\int \sec\chi' d\chi' = \ln|\sec\chi + \tan\chi|$.
More generally, because $2\alpha - 3 < -1$ for any $\alpha < 1$, this integral exhibits a non-integrable power divergence as $\chi \to \pi/2$, which drives the exponential suppression in $\Phi(t;t_A)$ to strictly zero, $\lim_{t \to t_B} \Phi(t; t_A) = 0$.
Combined with the exact relation $\mathbb{E}_{W}[(X(t) - \mu(t))^2] = \Phi^2(t; t_A) \tau(t; t_A) = \sigma^2(t)$, the variance strictly vanishes at the boundary, ensuring that the sample trajectories continuously satisfy the pinning condition $\lim_{t \to t_B} X(t) = \mu_B$ almost surely.

\subsection{Brownian bridge ($\alpha=1$)}
\label{sec:B_bridge}

Equation (\ref{eq:general_X_t_main}) reproduces the Brownian bridge 
under the matching condition $\alpha=1$.
Then the angular evolution equation (\ref{eqn:dif_eq_phi}) becomes $\sin \chi(t) = \omega (t - t_0)$ with a constant $t_0$.
The boundary conditions lead to $\chi(t_A) = -\pi/2$ and $\chi(t_B) = \pi/2$, and hence  we obtain 
\begin{equation}
\omega = \frac{2}{t_B - t_A} \, , \quad t_0 = \frac{t_A + t_B}{2} \, .
\end{equation}
Substituting into Eq.\ (\ref{eq:general_X_t_main}), the equation is simplified as 
\begin{align}
X(t) = \mu_A + \frac{\mu_B -\mu_A}{t_B -t_A} (t-t_A) 
+ \sqrt{2D} \int^t_{t_A} \frac{t_B - t}{t_B -s} \left( \frac{(t_B-t)(s-t_A)}{(t-t_A)(t_B -s)}\right)^{(\mathcal{K}_1-1)/2} dW_s \, .
\label{eqn:X_t_alpha=1}
\end{align}
This equation  yields the equation of the Brownian bridge by setting 
\begin{equation}
 \mathcal{K}_1 =   \frac{D}{R^2 \omega} = \frac{2D(t_B -t_A)}{(\mu_B -\mu_A)^2} = 1\, .
\end{equation}
This condition means that, unlike the standard Brownian bridge where the noise intensity $D$ and boundary values are independent \cite{revuz1999continuous}, here they are mutually constrained.
To distinguish this specific process, we refer to it as the canonical Brownian bridge \cite{koide_armin2025}.
The above constraints are required to reproduce the standard diffusive growth of variance in the Brownian bridge:
\begin{equation}
   \mathbb{E}_W [(X(t) - \mu_A)^2] = 2D (t-t_A) \, .
\end{equation}

So far, we have considered stochastic processes evolving over a given finite time interval ($t_A \le t \le t_B$). 
However, the type of stochastic process more familiar to us is one where only the initial time is fixed, and we consider the time evolution toward an arbitrary future time. 
Such a process corresponds to the limit where the time interval $t_B - t_A$ becomes very large in the context of a bridge process.

\subsection{Infinite limit of time interval}
\label{sec:infinite_limit}

In the following, we consider two distinct infinite limits of time interval 
in our generalized bridge process (\ref{eq:general_X_t_main}).

\subsubsection{Diffusion limit}
\label{sec:brownian_limit}

Let us consider the case of a general $\alpha$.
We first investigate the limit of $t_B - t_A \to \infty$ (or equivalently $\omega \to 0$) under the scaling condition that $D = R^2 \omega$ is held constant.
Since $R = (\mu_B - \mu_A)/2$ and $\omega \propto 2/(t_B - t_A)$, this operational limit physically corresponds to separating the final boundary state $\mu_B$ far away from $\mu_A$ proportionally to $\sqrt{t_B - t_A}$.

We consider a finite observation time $t \in (t_A, t_B)$ around a finite reference midpoint $t_0 = (t_A + t_B)/2$.
In this regime, the angular differential equation $\dot{\chi}(t) = \omega (\cos\chi)^{1-2\alpha}$ is evaluated near the top of the geodesic orbit ($\chi \approx 0$).
Because $\cos\chi \approx 1$ regardless of the value of $\alpha$, the angular velocity simplifies uniformly to $\dot{\chi} \approx \omega$, leading to the linear asymptotic form:
\begin{equation}
\chi(t) \approx \omega (t - t_0) \to 0 \, .
\end{equation}

Under the condition $D = R^2 \omega$ in the generalized bridge process (\ref{eq:general_X_t_main}),
the dimensionless coefficient $\mathcal{K}_\alpha$ remains a finite constant for any boundary-connecting geodesic ($\alpha > -1$): $\mathcal{K}_\alpha = 2/(1+\alpha)$.
Consequently, the integral inside the exponential function vanishes in the limit $\omega \to 0$ because the integration domain shrinks linearly with $\omega$.
Substituting these asymptotic relations into Eq.~(\ref{eq:general_X_t_main}), the generalized bridge process reduces directly to
\begin{equation}
X(t) = \mu_0 + \sqrt{2D} \int_{t_0}^t dW_s \, ,
\end{equation}
which is the exact solution to the diffusion SDE:
\begin{equation}
dX(t) = \sqrt{2D} dW_t \, .
\end{equation}

It is well-known that the canonical Brownian bridge converges to free Brownian motion in the long-time limit.
The analytical derivation above demonstrates that this convergence holds for all boundary-connecting geodesics ($\alpha > -1$): under the scaling $D = R^2 \omega$, the geometric drift induced by curvature or non-metricity vanishes, and the generalized bridge process reduces to free diffusion independently of the connection parameter $\alpha$.

\subsubsection{Ornstein-Uhlenbeck limit}
\label{sec:OU}

So far, we have realized the infinite limit by separating $\mu_B$ far away from $\mu_A$. 
We now consider the alternative physical regime: taking the same infinite limit of time interval while keeping both the spatial scale $R = (\mu_B - \mu_A)/2$ and the noise intensity $D$ fixed.

In this limit, our bridge process (\ref{eq:general_X_t_main}) spends the vast majority of its operation time near $\chi \approx 0$. 
By restricting our focus to the vicinity of this regime, the deterministic mean becomes effectively constant: $\mu(t) \approx \mu_0$.
Using the angular equation of motion, the time difference between two interior points can be expressed as
\begin{equation}
t - s = \int_{\chi(s)}^{\chi(t)} \frac{d\chi'}{\omega(\cos\chi')^{1-2\alpha}} \approx \frac{\chi(t) - \chi(s)}{\omega} \, .
\end{equation}
The integral inside the exponential factor in Eq.\ (\ref{eq:general_X_t_main}) simplifies directly to
\begin{equation}
\mathcal{K}_\alpha \int_{\chi(s)}^{\chi(t)} \frac{d\chi'}{(\cos\chi')^{3-2\alpha}} \approx \mathcal{K}_\alpha [\chi(t) - \chi(s)] := \gamma(\alpha)(t - s) \, ,
\end{equation}
where the relaxation rate $\gamma(\alpha)$ is a strictly positive constant for any boundary-connecting geodesic ($\alpha > -1$):
\begin{equation}
\gamma(\alpha) := \mathcal{K}_\alpha \omega = \frac{2D}{(1+\alpha)R^2} \, .
\end{equation}
Substituting these asymptotic expressions, the generalized bridge process reduces to
\begin{equation}
X(t) = \mu_0 + \sqrt{2D} \int_{t_A}^t e^{-\gamma(\alpha)(t - s)} dW_s \, , \label{eqn:X_OU_limit}
\end{equation}
which is a special solution to the standard Ornstein-Uhlenbeck (OU) equation:
\begin{equation}
dX(t) = -\gamma(\alpha) (X(t) - \mu_0) dt + \sqrt{2D} dW_t \, .
\label{eqn:OU_eq}
\end{equation}
Note that Eq.~(\ref{eqn:X_OU_limit}) corresponds to the solution with a constant mean $\mu(t) = \mu_0$.

It is instructive to highlight the following geometric property.
For the general solution of the OU equation starting from a deterministic initial state $(\mu(t_A), \sigma^2(t_A)) = (\mu_A, 0)$ with $|\mu_A - \mu_0| = R$, the time-dependent mean $\mu(t) = \mu_0 + (\mu_A - \mu_0)e^{-\gamma(\alpha)(t-t_A)}$ and variance $\sigma^2(t) = \frac{1+\alpha}{2}R^2 (1 - e^{-2\gamma(\alpha)(t-t_A)})$ trace out the exact same geometric ellipse of the Boundary-connecting case:
\begin{equation}
(\mu(t) - \mu_0)^2 + \frac{2}{1+\alpha} \sigma^2(t) = R^2 \, ,
\end{equation}
for all $t \ge t_A$.
Nevertheless, because the physical relaxation time $t$ governed by Eq.\ (\ref{eqn:OU_eq}) differs from the affine parameter of the geodesic equations (\ref{eqn:eta_geo_mu}) and (\ref{eqn:eta_geo_sigma}), the geodesic motion is represented not by the solution of Eq.\ (\ref{eqn:OU_eq}) but by Eq. (\ref{eq:general_X_t_main}).

Aside from the geometric classification of stochastic processes discussed above, our correspondence provides a practical, independent application: an operational method to determine the underlying connection parameter $\alpha$ for an observed non-equilibrium process.
In realistic laboratory experiments, continuous stochastic processes are rarely constrained as finite-time bridge processes.
Instead, they are predominantly observed as unconstrained relaxation processes toward equilibrium, governed by, for example, the standard OU equation.
If we interpret such an observed OU process (\ref{eqn:OU_eq}) as the asymptotic stationary limit of a geodesic, this relation allows us to identify the connection parameter $\alpha$ through macroscopic observables.
Specifically, by measuring the environmental noise intensity $D$, the physical relaxation rate $\gamma$, and the initial displacement scale $R = |\mu_A - \mu_0|$, 
the affine connection parameter is identified as
\begin{equation}
\alpha = \frac{2D}{\gamma R^2} - 1 \, .
\end{equation}
Rather than assuming an a priori fixed background geometry independent of the physical process, the statistical manifold in this operational framework is assigned to characterize a specific non-equilibrium relaxation process under a prescribed experimental setup.
In this operational view, the dependence of the connection on the initial preparation scale $R$ is not an inconsistency, but a natural consequence of using information geometry to encode non-equilibrium dissipative dynamics.

\subsection{Remarks on LC-geodesic in $(g, {}^{(0)}\Gamma)$ and Finite-Time Boundary Reachability}
\label{sec:LC_geodesic}

As discussed so far, when $\alpha > -1$, the geometric orbit of the geodesic forms an ellipse intersecting the deterministic boundary curve $\eta_2 = (\eta_1)^2$ (where $\sigma^2 = 0$) at two distinct points.
It is however easy to see that $\alpha$ should be positive for the time interval $\tau=t_B-t_A$ 
to be finite because  
\begin{equation}
\tau := t_B - t_A = \frac{1}{\omega} \int_{-\pi/2}^{\pi/2} (\cos\chi)^{2\alpha - 1} d\chi = \frac{\sqrt{\pi}}{\omega} \frac{\Gamma(\alpha)}{\Gamma\left(\alpha + \frac{1}{2}\right)} \, .
\label{eqn:total_time_gamma}
\end{equation}
Consequently, to formulate a bridge-like process over a finite operational time interval in the regime $-1 < \alpha \le 0$, one must introduce a finite-variance regulator $\epsilon^2 > 0$.
A prominent and physically important example of this situation is the Levi-Civita (LC) connection ($\alpha = 0$), where the Gaussian statistical manifold is strictly metric-compatible.
We then set the boundary conditions as $(\mu_A, \epsilon^2)$ at $t=t_A$ and $(\mu_B, \epsilon^2)$ at $t=t_B$.

Solving Eq.~(\ref{eqn:dif_eq_phi}) for $\alpha=0$ under the condition $\chi(t_0) = 0$, where $t_0 = (t_A + t_B)/2$ is the midpoint time, we find
\begin{equation}
\sec \chi + \tan\chi = e^{\omega(t - t_0)} \, ,
\end{equation}
which leads to 
\begin{align}
    \sec \chi &= \cosh (\omega(t - t_0)) \, ,\\
    \tan \chi &= \sinh (\omega(t - t_0)) \, .
\end{align}
Thus, the LC-geodesic is parameterized in time by
\begin{align}
\mu_{LC}(t) &= \mu_0 + R \tanh(\omega(t - t_0)) \, , \\
\sigma_{LC}^2(t) &= \frac{R^2}{2 \cosh^2(\omega(t - t_0))} \, ,
\end{align}
where the center $\mu_0$ and the characteristic radius $R$ are determined by the boundary conditions as 
\begin{align}
\mu_0 &= \frac{\mu_A + \mu_B}{2} \, ,\\
R^2 &=  \frac{(\mu_B - \mu_A)^2}{4} + 2\epsilon^2 \, .
\end{align}
From the analytical expression for $\sigma_{LC}^2(t)$ above, it is evident that it takes infinite physical time ($t \to \pm\infty$) for the trajectory to strictly intersect the deterministic boundary where $\sigma^2_{LC} = 0$.

Substituting these exact solutions into Eq.~(\ref{eqn:exact_sol_BridgeSDE}), we find that the LC-geodesics represent stochastic processes defined by: 
\begin{equation}
   X_{LC} (t) = \mu_{LC} (t) + \Phi_{LC} (t;t_A) W_{\tau_{LC}(t;t_A)} \, , \label{eqn:LC_bridge}
\end{equation}
where the propagator and time-scaling functions are evaluated as
\begin{align}
\Phi_{LC}(t;t_A) 
&= \frac{\cosh(\omega(t_A - t_0))}{\cosh(\omega(t - t_0))} \exp \left( -\frac{D}{2R^2\omega} \left[ 2\omega(t - t_A) + \sinh(2\omega(t-t_0)) - \sinh(2\omega(t_A-t_0)) \right] \right) \, , \\
\tau_{LC}(t; t_A) 
&= \frac{R^2}{4 \cosh^2(\omega(t_A - t_0))} \left\{ \exp\left( \frac{D}{R^2\omega} \left[ 2\omega(t - t_A) + \sinh(2\omega(t-t_0)) - \sinh(2\omega(t_A-t_0)) \right] \right) - 1 \right\} \, .
\end{align}

In this process, the angular velocity parameter $\omega$ is uniquely determined by the boundary conditions at $t=t_A$ and $t_B$ together with the infinitesimal regulator $\epsilon$:
\begin{equation}
\sigma^2_{LC}(t_A) = \sigma^2_{LC}(t_B) = \epsilon^2 > 0 \quad \implies \quad \omega = \frac{2}{t_B - t_A} \operatorname{arcosh} \left(\frac{R}{\sqrt{2}\epsilon}\right) \, . \label{eqn:omega_epsilon}
\end{equation}
Therefore, $\omega$ asymptotically vanishes in the long-time-interval limit ($t_B - t_A \to \infty$) whether with $D=R^2\omega$ constant or with $D$ and $R$ held fixed. 
Consequently, the subsequent asymptotic limits for this regularized LC-process proceed exactly as discussed before.

\subsection{Geometric Duality and Coordinate Transformation of Geodesics}
\label{sec:CooTrans}

In the preceding discussions, we formulated the general geodesics and their associated stochastic processes in terms of the expectation parameters $\eta$. 
The main objective of this section is to establish a direct mapping between the geodesics in the expectation parameters $\eta$ and those in the natural parameters $\theta$.

To establish this general mapping, let us consider the coordinate transformation from the expectation parameters $\eta_i$ to a new coordinate system $\tilde{\eta}^a$. 
Under this transformation, the dual connection coefficients transform as
\begin{equation}
{}^{(-\alpha)}\tensor{\Gamma}{^a_{bc}} (\tilde{\eta})
=
\left\langle d\tilde{\eta}^a , {}^{(-\alpha)}\nabla_{\frac{\partial}{\partial \tilde{\eta}^b}} \frac{\partial}{\partial \tilde{\eta}^c} \right\rangle
= \frac{\partial \tilde{\eta}^a}{\partial \eta_i} \frac{\partial \eta_j}{\partial \tilde{\eta}^b} \frac{\partial \eta_k}{\partial \tilde{\eta}^c} {}^{(-\alpha)}\tensor{\Gamma}{_i^{jk}} (\eta)   
+ \frac{\partial \tilde{\eta}^a}{\partial \eta_m} \frac{\partial^2 \eta_m}{\partial \tilde{\eta}^b \partial \tilde{\eta}^c} \, ,
\end{equation}
where $\langle \omega, X \rangle = \omega(X)$ denotes the natural pairing between a $1$-form $\omega$ and a vector field $X$.
When we choose the natural parameters as our new coordinates ($\tilde{\eta}^a = \theta^a$) through the Legendre transformation, using $\partial \theta^a / \partial \eta_i = g^{ai}(\eta)$ and $\partial \eta_j / \partial \theta^b = g_{jb}(\theta)$, the explicit evaluation of the right-hand side yields the transformation from ${}^{(-\alpha)}\tensor{\Gamma}{_i^{jk}} (\eta)$ to ${}^{(-\alpha)}\tensor{\Gamma}{^a_{bc}} (\tilde{\eta})$ as
\begin{equation}
{}^{(-\alpha)}\tensor{\Gamma}{_i^{jk}} (\eta) \longrightarrow \,\,
{}^{(-\alpha)}\tensor{\Gamma}{^a_{bc}} (\tilde{\eta}) = {}^{(-\alpha)}\tensor{\Gamma}{^a_{bc}} (\theta) \label{eqn:flat_transformation} \, .
\end{equation}
Equation (\ref{eqn:flat_transformation}) demonstrates that the geodesics governed by ${}^{(-\alpha)}\Gamma (\eta)$ in the expectation-parameter space are transformed to the geodesics governed by ${}^{(-\alpha)}\Gamma (\theta)$ in the natural-parameter space.

Therefore, the expectation parameter geodesics on the manifold equipped with $(g, {}^{(\alpha)}\Gamma, {}^{(-\alpha)}\Gamma)$ correspond directly to the natural parameter geodesics on the manifold equipped with the inverted connection $(g, {}^{(-\alpha)}\Gamma, {}^{(\alpha)}\Gamma)$, where the first connection is associated with the natural parameters and the second with the expectation parameters.
Since we have already solved the geodesics of the expectation parameters for all $\alpha \in \mathbb{R}$ in the preceding sections, the geodesics of the natural parameters for all $\alpha \in \mathbb{R}$ are simultaneously identified.

\section{Geodesic interpretation of thermodynamically optimal protocol}
\label{sec:optimization}

So far, we have studied the relation between stochastic processes and geodesics of the Gaussian statistical manifold.
In this section, we show that 
the geodesics with non-metricty of the Gaussian statistical manifold can give the optimal protocol in stochastic thermodynamics.
Let us consider a one-dimensional overdamped Fokker-Planck equation where a Brownian particle is confined within a harmonic potential $V(x, a_t) = a_t x^2/2$, and $a_t$ is an external parameter with which the width of the potential is controlled.
We consider an initial thermal equilibrium state with $a_{A}$ and change the width to $a_{B}$.
In the limit of infinite operation time $\tau = t_B -t_A$, 
the work necessary to realize this process is equivalent to the difference in the Helmholtz free energies as is well known in thermodynamics. 
When the operation time $\tau$ is fixed to a finite value, this process becomes irreversible.
The optimal protocol minimizes this irreversible work \cite{sasa2001,sekimoto_book}.
In this optimization problem, there are two distinct perspectives for the variational calculation: one fixes the initial and final values of the potential width $a_t$ \cite{sasa2001,sekimoto_book,sivak2012,koide2017}, while the other fixes the initial and final probability distributions (or statistical moments) \cite{Schmiedl2007,aurell2011,Zhong2024}. 
Because the instantaneous probability distribution depends on the history of the protocol $a_t$, these two approaches generally lead to different variational problems. 
In the former approach, the exact optimal protocol which minimizes the irreversible work is given by the solution of the following integro-differential equation \cite{koide2017}:
\begin{align}
& \int^t_{t_A} ds \frac{\dot{a}_s}{a^2_s} \partial_t e^{-2\beta D \int^t_s du a_u} + \frac{1}{a^2_t} \int^{t_B}_t ds \dot{a}_s \partial_t e^{-2\beta D \int^t_s du a_u} \nonumber \\
& + 2\beta D \int^{t_B}_t ds_2 \int^t_{t_A} ds_1 e^{-2\beta D\int^{s_2}_{s_1}dy a_u } \frac{\dot{a}_{s_1}\dot{a}_{s_2}}{a^2_{s_1}} =0 \, ,
\end{align}
where $D$ is a diffusion constant and $\beta$ denotes the inverse temperature of the heat bath.

When the time scale of the parameter variation is sufficiently slow and the nonequilibrium Gaussian distribution remains close to the corresponding equilibrium distribution determined by $a_t$, the optimal protocols obtained from both perspectives coincide with each other. 
For such a long operation time $\tau$, this complex equation reduces to the well-known result \cite{sasa2001,sekimoto_book,Schmiedl2007}:
\begin{equation}
\ddot{a}_t - \frac{3}{2a_t} \dot{a}_t^2 = 0 \, , 
\label{eqn:simplified_optimized_protocol_eq}
\end{equation}
leading to 
\begin{equation}
    a_t = \frac{(t_B - t_A)^2 a_A a_B}{ \{ (t-t_A)\sqrt{a_A} + (t_B - t)\sqrt{a_B} \}^2} \, .
\end{equation}
It is known that the behavior of the optimal protocol in the slow-driving limit can be reproduced using thermodynamic geometry \cite{sivak2012} or the optimal transport problem based on Wasserstein geometry \cite{aurell2011,Zhong2024}. 
The former approach relates to our discussion, but the metric is generally modified from the Fisher metric and does not account for non-metricity. 
Wasserstein geometry defines a metric on the space of probability distributions through optimal transport theory, where the distance is induced by the cost of physically transporting material from one distribution to another. 
Here, we show that an alternative geometric perspective on this optimal protocol is possible: namely, the optimal protocol in the slow-driving limit can be interpreted as an expectation geodesic on the Gaussian statistical manifold endowed with a non-metric connection.

The instantaneous equilibrium distribution is a Gaussian distribution, which is the ground state of the time-dependent Fokker-Planck operator. 
According to the perturbative expansion with respect to the control speed $\dot{a}_t$ \cite{koide2017}, the nonequilibrium distribution $\rho(x, t)$ is represented by a superposition of the instantaneous equilibrium distribution $\rho_{eq}(x, a_t)$ and the excited states. 
In the linear response regime, the dominant correction comes from the second excited state, which contains the Hermite polynomial of degree two (proportional to $x^2$). 
Therefore, the nonequilibrium distribution is approximately given by
\begin{equation}
\rho(x, t) \approx \rho_{eq}(x, a_t) \left[ 1 + \epsilon(t) x^2 + C(t) \right] \, ,
\end{equation}
where $\epsilon(t)$ is a small quantity proportional to $\dot{a}_t$, and $C(t)$ is a normalization constant. 
Since $\epsilon(t)$ is small in the limit of infinite operation time, 
we can re-exponentiate the term inside the bracket as $1 + \epsilon(t) x^2 \approx \exp(\epsilon(t) x^2)$. 
By substituting the explicit form of the equilibrium distribution $\rho_{eq} \propto \exp(-\beta a_t x^2 / 2)$, we obtain
\begin{equation}
\rho(x, t) \approx \exp \left( -\frac{x^2}{2\sigma^2} \right) \, ,
\end{equation}
where
\begin{equation}
    \sigma^2 = \frac{1}{\beta a_t - 2 \epsilon(t)} \, .
\end{equation}

In the present case ($\dot{\mu}=0$), the expectation geodesic equation for the variance $\sigma^2$, related to the second expectation parameter through $\eta_2 = \mu^2 + \sigma^2$, is governed by Eq.~(\ref{eqn:eta_geo_sigma}), yielding
\begin{equation}
\frac{d^2 \sigma^2}{dt^2} - \frac{1-\alpha}{\sigma^2} \left( \frac{d \sigma^2}{dt}  \right)^2 = 0 \, .
\end{equation}
In the limit of infinite operation time where $\epsilon(t) = 0$, 
the substitution of this relation into the geodesic equation yields
\begin{equation}
\ddot{a}_t - \frac{1+\alpha}{a_t} \dot{a}_t^2 = 0 \, .
\end{equation}

By setting $\alpha=1/2$, this equation reduces to the equation of the optimal protocol in the slow-driving limit (\ref{eqn:simplified_optimized_protocol_eq}). 
That is, the simplified optimal protocol mathematically corresponds to the geodesic of expectation parameters in the Gaussian statistical manifold equipped with $(g,{}^{(1/2)}\Gamma,{}^{(-1/2)}\Gamma)$. 
This result indicates that non-metricity, which is typically neglected in standard applications of information geometry to physics, can play an important role in capturing the geometric structure of optimal thermodynamic protocols.
Note that, in practical thermodynamic protocols, the physical operation takes place between initial and final states with strictly finite positive variances ($\sigma_A^2, \sigma_B^2 > 0$).
Thus, the optimal protocol corresponds to a smooth, regular segment of the $\alpha = 1/2$ expectation geodesic, entirely avoiding the singular deterministic boundary ($\sigma^2 = 0$) where the required trap stiffness would diverge ($a_t \to \infty$).

\section{The Gravitational Aspect of Information}
\label{sec:gravity}

We have investigated the properties of geodesics on the Gaussian statistical manifold and found that motion along a geodesic admits physical significance through associated stochastic processes.
We can understand this significance by analogy with general relativity \cite{koide_armin2025,koide_armin_review,wada2026}.
A stochastic process evolving along a geodesic can be interpreted as a ``free" motion in the sense that it represents a time evolution uninfluenced by external or artificial constraints on the statistical manifold.  
In information geometry, however, the physical interpretation of this ``free" motion strictly depends on the chosen affine connection, which serves as the underlying geometric representation.
For instance, if we equip the Gaussian statistical manifold with the metric-compatible LC connection ($\alpha = 0$), the space is curved.
In this representation, the ``free" geodesic motion is governed by geometric curvature, which acts as a restoring force that naturally forces the process to reduce to the OU process in the infinite time-interval limit.  
Conversely, if we select a specific flat connection ($\alpha = 1$), the manifold becomes completely flat.  
In this distinct geometric picture, the ``free" motion lacks any restoring force and corresponds to the canonical Brownian bridge.
Furthermore, for other non-metric connections ($\alpha \neq 1$), the geometric drift induced by non-metricity acts as a restoring force, again reproducing the OU process.
This demonstrates that our physical interpretation of a given stochastic process depends heavily on the chosen geometric representation of the manifold.

Interestingly, this representation-dependent interpretation directly reflects a fundamental concept in modern gravitational physics: the Geometrical Trinity of Gravity \cite{trinity,iosifidis2023,koide_armin_review,wada2026}.
In standard general relativity, gravitational effects are attributed entirely to the curvature of spacetime.
However, mathematically equivalent reformulations demonstrate that the exact same gravitational phenomena can be described on a completely flat spacetime by attributing them either to torsion (Teleparallel Equivalent of General Relativity, TEGR) or to non-metricity (Symmetric Teleparallel Equivalent of General Relativity, STEGR).
There exists a clear conceptual parallel between this gravitational trinity and our information-geometric findings.
The restoring force observed in a stochastic process, such as the OU process, corresponds to a physical drift.
We can geometrically interpret this force as originating from curvature when adopting the metric-compatible LC representation ($\alpha = 0$).
Conversely, when using a flat non-metric connection, we interpret the exact same restoring force as originating intrinsically from non-metricity.
This conceptual analogy indicates that exploring the choice of connections in information geometry provides a valuable mathematical laboratory \cite{iosifidis2023,wada2026} to deepen our understanding of the physical implications of non-metricity and alternative geometric representations in Metric-Affine gravity \cite{trinity,koide_armin_review}.

\section{Concluding remarks}
\label{sec:conclusion}

In this study, we established a systematic correspondence between $\alpha$-geodesics on the Gaussian statistical manifold and continuous stochastic processes within the framework of information geometry.
Focusing on the Boundary-connecting class of geodesics ($\alpha > -1$) that intersect the deterministic boundary ($\sigma^2 = 0$) at two distinct times,
we formulated the generalized bridge process, with the canonical Brownian bridge serving as the standard geometric trajectory for $\alpha = 1$.
Under appropriate operational limits, this generalized bridge process reduces to fundamental stochastic processes: either Ornstein-Uhlenbeck (OU) relaxation or free diffusion.
In the class of linear Gauss-Markov dynamics driven by time-independent additive Brownian noise, this formulation demonstrates how continuous stochastic processes can be systematically classified and extended along geometric curves.

From a broader mathematical perspective, however, a curve of one-time marginal distributions on the statistical manifold can be lifted to other stochastic dynamics by allowing time-dependent noise amplitudes or incorporating memory effects, such as non-Markovian stochastic differential equations.
Investigating whether non-metric affine connections can systematically classify and characterize these non-Markovian dynamics and generalized diffusion processes remains an open question for future geometric physics.


Furthermore, we demonstrated that the optimal protocol for minimizing irreversible work in the slow-driving limit exactly coincides with the expectation geodesic associated with a specific non-metric connection with ${}^{(-1/2)}\Gamma(\eta)$.
The coincidence between the optimal protocol and the geodesic of expectation parameters, however, does not necessarily imply that the Gaussian statistical manifold equipped with $(g,{}^{(1/2)}\Gamma,{}^{(-1/2)}\Gamma)$ provides the most appropriate geometric description for the optimal time evolution of this one-dimensional overdamped system with a harmonic potential.
Indeed, when we apply information geometry to actual physical systems, we can consider a wide variety of distinct perspectives regarding the role of geodesics.
To illustrate this diversity, let us consider two possible approaches from among these various perspectives.
First, one can adopt the view that the unforced, natural time evolution of a system should trace a geodesic.
Under this framework, an optimal protocol under external control is understood as a controlled trajectory determined by minimizing a specific cost function that quantifies the deviation from this unforced, natural geodesic.
For instance, when controlling a diffusion process through an external force, it is physically more natural to select the flat statistical manifold equipped with $(g,{}^{(1)}\Gamma,{}^{(-1)}\Gamma)$, where the free diffusion itself corresponds to an expectation geodesic.
As another alternative, one can propose that the connection parameter $\alpha$ should be directly determined from macroscopic physical observables.
As we demonstrated, the expectation geodesic asymptotes to the OU process in the infinite limit of the time interval, described by the relaxation rate $\gamma(\alpha) = \frac{2D}{(1+\alpha)R^2}$.
This relation allows us to identify the manifold parameter $\alpha$ from the environmental noise intensity $D$, the observed relaxation rate $\gamma$, and the initial displacement scale $R=|\mu_A - \mu_0|$.
As these different perspectives illustrate, the choice of the appropriate statistical manifold depends intimately on whether one prioritizes the unforced diffusive background or the observed dissipative dynamics.
Establishing comprehensive physical criteria to select and interpret the underlying affine connections across diverse non-equilibrium systems remains an important perspective for future study.

We also highlighted a conceptual analogy between our information-geometric representation of stochastic processes and the Geometrical Trinity of Gravity.
Just as gravitational phenomena can be equivalently described through curvature, torsion, or non-metricity, the same physical restoring force in stochastic dynamics can be attributed either to curvature or to non-metricity depending on the chosen affine connection.
Because standard information geometry restricts itself to torsionless affine connections, our correspondence specifically mirrors the duality between curvature and non-metricity.
Investigating how this geometric correspondence extends when incorporating torsion through asymmetric connections on statistical manifolds remains an intriguing direction for future study.

Recently, Ito proposed a unified geometric thermodynamics for the Fokker-Planck equation by linking information geometry and optimal transport theory \cite{Ito2024}.
While his work primarily focuses on the Riemannian metric structures and thermodynamic costs associated with optimal transport, our research highlights the physical role of the non-metric connection.
In this respect, our work and Ito's framework present a complementary and contrasting picture, and this comparison demonstrates that both metric and connection are essential to fully grasp the geometric nature of stochastic systems.

It is of great interest to compare our results with the findings of Heseltine and Kim \cite{Heseltine2019}.
Their work investigates which metric most appropriately describes the relaxation trajectory of the OU process in information space.
They conclude that only the quantity called information length exhibits a unique linear dependence on the initial mean position.
This linear geometric property of the relaxation process might be deeply connected to our result, where the expectation geodesics strictly reproduce the OU process only in the infinite time-interval limit.

The correspondence between $\alpha$-geodesics and stochastic processes established in this work extends its relevance to modern generative AI.
Recent literature highlights that the geometry of the latent space in diffusion models can be successfully clarified through the reverse SDE and the Fisher metric \cite{Karczewski2025}.
While conventional machine learning applications often restrict themselves to standard metric-compatible or specific Riemannian manifolds, our generalization to arbitrary non-metric connections introduces a new degree of freedom to characterize and optimize generative trajectories. The application of our framework to geometric deep learning remains a promising future task.

As a final remark, these results provide an important foundation to apply information geometry to quantum systems in the future \cite{petz1996,holevo,helstrom,Paris2009,koide_armin_tmography}.
To successfully develop quantum information geometry, the theoretical framework must rigorously reproduce the properties of classical information geometry in the classical limit.
As a prominent recent example, a similar geodesic interpretation successfully minimizes nonadiabatic entropy production in transitions between quantum nonequilibrium steady states \cite{Scandi2019,Abiuso2020,Rolandi2023,Lacerda2025-1,Lacerda2025-2}.

\begin{acknowledgments}
The authors thank L. Bettmann for useful discussions related to Ref.\ \cite{Lacerda2025-2}, which significantly contributed to the development of Sec.\ \ref{sec:optimization}.
T.K. acknowledges the financial support by CNPq (No.\ 304504/2024-6). A.vdV. gratefully acknowledges funding by the Deutsche Forschungsgemeinschaft (DFG, German Research Foundation) -- Project number 570900169.
A part of this work has been done under the project INCT-Nuclear Physics and Applications (No.\ 408419/2024-5.).
\end{acknowledgments}

\appendix

\section{Finite Convergence of $\Phi(t;t_A)$ at the Deterministic Boundary}
\label{app:finite}

In Sec.~\ref{sec:sto_process_GGM}, we introduced the function $\Phi(t; t_A)$ to analyze the stochastic dynamics near the deterministic boundary $t = t_A$, where the variance vanishes ($\sigma^2(t_A) = 0$). 
Here, we confirm that $\Phi(t; t_A)$ converges to a finite non-zero value as $t \to t_A$.

For continuous diffusion processes with constant noise intensity $D$ starting from a deterministic state $\sigma^2(t_A) = 0$, the initial variance growth is dominated by Brownian diffusion.
Near the boundary $t \approx t_A$, the variance admits the general asymptotic expansion:
\begin{equation}
\sigma^2(t) = 2D(t - t_A) + c_2 (t - t_A)^2 + \mathcal{O}((t - t_A)^3) \, ,
\label{eqn:sigma2_boundary_expansion}
\end{equation}
where $c_2$ is a system-dependent expansion coefficient.
Indeed, evaluating our exact geodesic solution (\ref{eqn:geo_evolution_sigma2}) near $\chi(t_A) = -\pi/2$ directly yields this linear initial growth with $\left. d\sigma^2/dt \right|_{t=t_A} = 2D$.
Consequently, the standard deviation behaves asymptotically as $\sigma(t) = \sqrt{2D(t - t_A)} \left( 1 + \frac{c_2}{4D}(t - t_A) + \mathcal{O}((t - t_A)^2) \right)$.

We examine the exponential factor in the representation
\begin{equation}
\Phi(t; t_{\mathrm{ref}}) = \frac{\sigma(t)}{\sigma(t_{\mathrm{ref}})} \exp \left( -\int_{t_{\mathrm{ref}}}^t \frac{D}{\sigma^2(u)} du \right) \, ,
\end{equation}
where $t_{\mathrm{ref}} \in (t_A, t_B)$ is a reference interior time.
Using Eq.~(\ref{eqn:sigma2_boundary_expansion}), the integrand near the boundary is expanded as
\begin{equation}
\frac{D}{\sigma^2(u)} = \frac{1}{2(u - t_A)} - \frac{c_2}{4D} + \mathcal{O}(u - t_A) \, .
\end{equation}
Integration from $t_{\mathrm{ref}}$ to $t$ (for $t_A < t < t_{\mathrm{ref}}$) yields
\begin{equation}
\exp \left( - \int_{t_{\mathrm{ref}}}^t \frac{D}{\sigma^2(u)} du \right) 
= \frac{\sqrt{t_{\mathrm{ref}} - t_A}}{\sqrt{t - t_A}} \exp \left( \frac{c_2}{4D}(t - t_{\mathrm{ref}}) \right) \left[ 1 + \mathcal{O}(t - t_A) \right] \, .
\end{equation}
While this exponential factor diverges as $1/\sqrt{t - t_A}$ in the limit $t \to t_A$, the prefactor $\sigma(t)$ simultaneously approaches zero as $\sqrt{t - t_A}$.
The singular algebraic factor $\sqrt{t - t_A}$ cancels out exactly, leading to the finite boundary limit:
\begin{equation}
\lim_{t \to t_A} \Phi(t; t_{\mathrm{ref}}) = \frac{\sqrt{2D(t_{\mathrm{ref}} - t_A)}}{\sigma(t_{\mathrm{ref}})} \exp \left( \frac{c_2}{4D}(t_A - t_{\mathrm{ref}}) \right) \, .
\end{equation}
Setting $t_{\mathrm{ref}} \to t_A$ ensures that $\Phi(t_A; t_A) = 1$.
Therefore, $\Phi(t; t_A)$ remains finite and well-defined on the boundary $t=t_A$, despite the apparent divergence in the exponent.



\begin{thebibliography}{99}
%
\bibitem{amari1985}
S.\ Amari,
\textit{Differential-Geometrical Methods in Statistics},
(Springer-Verlag, Berlin, 1985).
%
\bibitem{amari2000}
S.\ Amari and H.\ Nagaoka, 
\textit{Methods of Information Geometry}, 
(Oxford University Press, and American Mathematical Society, Providence, 2000).
%
\bibitem{amari2016}
S.\ Amari,
\textit{Information Geometry and its Applications},
(Springer, Japan, 2016).
%
\bibitem{ay2017}
N.\ Ay, J.\ Jost, H.\ V.\ L\^{e}, and L.\ Schwachh\"{o}fer,
\textit{Information Geometry},
(Springer, Switzerland, 2017).
%
\bibitem{brody2009}
D.\ C.\ Brody and D.\ W.\ Hook,
``Information geometry in vapour-liquid equilibrium",
J. Phys. A, \textbf{42}, 023001 (2009).
%
\bibitem{ItoDechant2020} 
S.\ Ito and A.\ Dechant, ``Stochastic Time Evolution, Information Geometry, and the Cram\'er-Rao Bound", Phys. Rev. X \textbf{10}, 021056 (2020).
%
\bibitem{Kim2021}
E.-j.\ Kim, ``Information Geometry, Fluctuations, Non-Equilibrium Thermodynamics, and Geodesics in Complex Systems,'' 
Entropy, \textbf{23}, 1393 (2021).
%
\bibitem{Ito2024}
S.\ Ito, 
``Geometric thermodynamics for the Fokker-Planck equation: stochastic thermodynamic links between information geometry and optimal transport", 
Information Geometry, \textbf{7}, 441 (2024).
%
\bibitem{weyl1918english}
H.\ A.\ Lorentz, A. Einstein, H.\ Minkowski and H.\ Weyl, \textit{The Principle of Relativity A Collection of Original Memoirs on the Special and General Theory of Relativity} (Dover, New York, 1923).
%
\bibitem{Hehl95}
F.\ W.\ Hehl and J.\ D.\  McCrea, and E.\ W.\  Mielke, and Y.\ Ne'eman, 
``Metric affine gauge theory of gravity: Field equations, Noether identities, world spinors, and breaking of dilation invariance",
Phys. Rept \textbf{258}, 1 (1995).
%
\bibitem{Hehl99}
F.\ W.\ Hehl and A.\ Macias, 
``Metric affine gauge theory of gravity. 2. Exact solutions",
Int.\ J.\ Mod.\ Phys.\ D \textbf{8}, 399 (1999).
%
\bibitem{Baldazzi21}
A.\ Baldazzi, O.\  Melichev and R.\ Percacci,
``Metric-Affine Gravity as an effective field theory",
Ann.\ Phys. \textbf{438}, 168757 (2022).
%
\bibitem{Francois25}
J.\ Fran{\c{c}}ois and L.\ Ravera, 
``Reassessing the foundations of metric-affine gravity",
Eur.\ Phys.\ J.\ C \textbf{85}, 902 (2025).
%
\bibitem{Blagojevic12}
M.\ Blagojevic and F.\ W.\ Hehl,
``Gauge Theories of Gravitation", 
arXiv:1210.3775 (2012).
%
\bibitem{Obukhov06}
Y.\ N.\ Obukhov,
``Poincare gauge gravity: Selected topics",
Int.\ J.\ Geom.\ Meth.\ Mod.\ Phys. \textbf{3}, 95 (2006).
%
\bibitem{Mielke17}
E.\ W.\ Mielke, 
{\it Geometrodynamics of Gauge Fields. On the Geometry of Yang-Mills and Gravitational Gauge Theories},
(Springer, Berlin, 2017)

%
\bibitem{Karczewski2025}
R.\ Karczewski, M.\ Heinonen, A.\ Pouplin, S.\ Hauberg and V.\ Garg,
``The Spacetime of Diffusion Models: An Information Geometry Perspective,'' 
arXiv:2505.17517 .
%
\bibitem{ohara}
A.\ Ohara,
``Geometric study for the Legendre duality of generalized entropies and its application to the porous medium equation,"
Eur. Phys. J. \textbf{B70}, 15 (2009).
%
\bibitem{oharawada2010}
A.\ Ohara and T.\ Wada,
``Information geometry of $q$-Gaussian densities and behaviors of solutions to related diffusion equations,''
J. Phys. A: Math. Theor. \textbf{43}, 035002 (2010).
%
\bibitem{trinity}
J.\ B.\ Jim\'{e}nez, L.\ Heisenberg and T.\ S.\ Koivisto, ``The Geometrical Trinity of Gravity", Universe \textbf{5}, 173 (2019).
%
\bibitem{iosifidis2023}
D.\ Iosifidis and K.\ Pallikaris, 
``Biconnection gravity as a statistical manifold", 
Phys. Rev. D \textbf{108}, 044026 (2023).
%
\bibitem{wada2026}
T. Wada and A. M. Scarfone, ``Non-Metricity in Information Geometry,'' 
Entropy \textbf{28}, 447, (2026).
%
\bibitem{koide_armin_review}
T.\ Koide and A.\ van de Venn,
``Torsion-Induced Quantum Fluctuations in Metric-Affine Gravity
Using the Stochastic Variational Method",
Symmetry \textbf{18}, 525 (2026).
%
\bibitem{revuz1999continuous}
D.\ Revuz and M.\ Yor, 
\textit{Continuous Martingales and Brownian Motion}
(Springer, New York, 1999).
%
\bibitem{Gardiner_book}
C.\ W.\ Gardiner \textit{Handbook of Stochastic Method: for Physics, Chemistry and Natural Sciences}, 
(Springer, New York, 2004)
%
\bibitem{koide_armin2025}
T.\ Koide and A.\ van de Venn, 
``The Gravitational Aspect of Information:
The Physical Reality of Asymmetric “Distance”, 
arXiv:2510.22664.
%
\bibitem{wada2026}
T.\ Wada, ``REFRAMING OF INFORMATION GEOMETRY VIA SYMMETRIC
TELEPARALLEL GRAVITY",
arXiv:2607.00649.
%
\bibitem{sasa2001}
K.\ Sekimoto and S.\ Sasa,
``Complementarity relation for irreversible process derived from
stochastic energetics",
J. Phys. Soc. Japan \textbf{66}, 3658 (2001).
%
\bibitem{sekimoto_book}
K.\ Sekimoto, \textit{Stochastic Energetics} (Berlin, Springer, 2010).
%
%
%
\bibitem{koide2017}
T.\ Koide, 
``Perturbative expansion of irreversible work in Fokker-Planck equation \'{a} la quantum mechanics",
J. Phys. A: Math. Theor. \textbf{50}, 325001 (2017).
%
\bibitem{sivak2012}
D.\ A.\ Sivak and G.\ E.\ Crooks, 
``Thermodynamic metrics and optimal paths",
Phys. Rev. Lett. \textbf{108}, 190602 (2012).
%
\bibitem{Schmiedl2007}
M.\ R.\ Schmiedl and U.\ Seifert, 
``Optimal finite-time processes in stochastic thermodynamics", 
Phys. Rev. Lett. \textbf{98} 108301 (2007).
%
%
\bibitem{aurell2011}
E.\ Aurell, C.\ Mej\'{i}a-Monasterio, and P.\ Muratore-Ginanneschi, 
``Optimal protocols and optimal transport in stochastic thermodynamics",
Phys. Rev. Lett. \textbf{106}, 250601 (2011).
%
\bibitem{Zhong2024}
A.\ Zhong and M.\ R.\ DeWeese, 
``Beyond linear response: Equivalence between thermodynamic geometry and optimal transport", 
Physical Review Letters, \textbf{133}, 057102 (2024).
%
\bibitem{Heseltine2019}
J.\ Heseltine and E.-j.\ Kim, 
``Comparing information metrics for a coupled Ornstein-Uhlenbeck process", 
Entropy, \textbf{21}, 775 (2019).
%
\bibitem{helstrom}
C. W. Helstrom, \textit{Quantum Detection and Estimation Theory}, (Academic Press, London, 1976).
%
\bibitem{petz1996}
D. Petz,
``Monotone metrics on matrix spaces,"
\textit{Linear Algebra Appl.} \textbf{244}, 81 (1996).
%
\bibitem{Paris2009}
M.\ G.\ A.\ Paris, 
``Quantum estimation for quantum technology", 
Int. J. Quant. Inf. \textbf{7}, 125 (2009).
%
\bibitem{holevo}
A. S. Holevo, 
\textit{Probabilistic and Statistical Aspects of Quantum Theory}, (Edizioni della Normale, 2011).
%
\bibitem{koide_armin_tmography}
T.\ Koide and A. van de Venn, 
``Information-Geometric Quantum Process Tomography of Single
Qubit Systems",
Phys. Lett. A \textbf{594}, 132074 (2026).
%
\bibitem{Scandi2019}
M.\ Scandi and M. Perarnau-Llobet,
``Thermodynamic length in open quantum systems",
Quantum \textbf{3}, 197 (2019).
%
\bibitem{Abiuso2020}
P.\ Abiuso,H.\ J.\ D.\ Miller,M.\ Perarnau-Llobet and M.\ Scandi,
``Geometric Optimisation of Quantum Thermodynamic Processes",
Entropy \textbf{22}, 1076 (2020).
%
\bibitem{Rolandi2023}
A.\ Rolandi and M.\ Perarnau-Llobet,
``Finite-time Landauer principle beyond weak coupling",
Quantum \textbf{7}, 1161 (2023).
%
\bibitem{Lacerda2025-1}
L.\ P.\ Bettmann and J.\ Goold,
``Information geometry approach to quantum stochastic thermodynamics",
Phys.\ Rev. E \textbf{111}, 014133 (2025).
%
\bibitem{Lacerda2025-2}
A.\ M.\ Lacerda, L.\ P.\ Bettmann and J.\ Goold,
``Information geometry of transitions between quantum nonequilibrium steady states",
Phys. Rev. E \textbf{112}, L022101 (2025).
\end{thebibliography}
\end{document}